\documentclass[12pt]{article}
\usepackage[utf8]{inputenc}
\usepackage{graphicx}
\usepackage{hyperref}

\begin{document}
\title{Musical Excellence of Mridangam by Dr. Umayalpuram K Sivaraman, Dr. T Ramasami and Dr. M D Naresh}
\author{An introductory review by Dr. Arvind Shankar Kumar}
\maketitle
\newpage
\topskip0pt
\vspace*{\fill}
\textit{To Sivaraman Sir, whose evergreen enthusiasm and quest for excellence continues to inspire and astound in equal measure.}
\vspace*{\fill}

\newpage
\tableofcontents
\newpage
\section{Preface}

Musical Excellence of Mridangam is a treatise by the legendary mridangam artiste Padma Vibhushan awardee Sangeetha Kalanidhi Dr. Umayalpuram K Sivaraman in collaboration with eminent scientists Dr. Ramasami and Dr. Naresh of Central Leather Research Institute Chennai, whose grand ambition is to redefine our understanding of the ancient instrument that is the mridangam, using both modern scientific tools and artistic insight. One of the most challenging aspects of such an ambitious undertaking is to interpret the subjective experience of music using tools designed for objective analysis. Given this paradox, there are bound to be gaps in our interpretation and understanding that cannot be bridged. However, as Musical Excellence and previous undertakings of such nature - mostly confined to Western music - have repeatedly shown us, there is tremendous insight to be gained through such an analysis. However, such scientific analysis is quite sparse in the field on Carnatic music, whose beauty and uniqueness still remains shrouded in myth and mystique. In the case of the mridangam, the famed physicist Dr. CV Raman was the first to pursue this avenue, and he made important discoveries in the 1920s regarding the unique nature of the mridangam within the world of musical instruments, which have been carried on by researchers in the following decades. Musical Excellence builds on this and delivers a treatise which comprehensively analyses the unique tonal properties of the mridangam, unravels the role of various materials and traditional fabrication methodologies, and proposes new designs for travel-friendly mridangams and methods of standardization of materials and manufacturing which preserve the musical experience produced.

The primary value of such an enterprise as Musical Excellence of Mridangam is expected to be 1) to the rasika, who can gain a new appreciation which enhances their experience of music, 2) to the artist, who can use the enhanced understanding of their instrument to develop new techniques to produce musical experiences and finally 3) to the makers of the instruments, who can use the understanding of the role of various materials, designs and fabrication processes to improve and innovate new designs to enhance the quality of musical notes produced. One aspect that we note about these three categories, which is not typical of the audience of any other kind of scientific treatise, is that it does not include trained scientists. However, the rigour of scientific analysis requires that new discoveries and insights are presented in a language that reflects said rigour and understanding of the analysis techniques used, and thus may not be directly accessible to people in any of these categories. 

This review thus aims to bridge the gap between the primary intended audience of Musical Excellence of Mridangam and the scientific rigour with which the original treatise is written by introducing the subject of musical analysis and presenting the discoveries made about the wonders of this unique instrument in Musical Excellence to rasikas, artistes and makers alike. The first three chapters of this review (including this one) introduces the basic scientific concepts used in Musical Excellence of Mridangam and provides background to previous scientific research into this instrument, starting from the seminal work of Dr. CV Raman. This also includes brief discussions of the corresponding chapters in Musical Excellence of Mridangam. The next chapters all serve the purpose of explaining the main scientific results presented in Musical Excellence of Mridangam in each of the corresponding chapters in the treatise, and finally summarizing the relevance of the work.

\newpage

\section{Introduction to Musical Analysis}

\subsection{What makes a musical note? - A primer on harmonics}
Sound is nothing but variations of pressure in the atmosphere, that are picked up by the sensitive instruments that are our eardrums. However, certain sounds are pleasant to hear - which we call 'musical' - and certain sounds unpleasant - which we call 'noise'. Our ears are therefore able to make qualitative distinctions between different kinds of pressure variations in the atmosphere. In this chapter, we will see that this qualitative distinction has a basis in the theory of waves and our ears are, in a sense, a form of spectral analysers (a sophisticated scientific instrument found in electronics labs).
\subsubsection{Visualizing sound}
\begin{figure}[htb!]
    \centering
    \includegraphics[width=10cm]{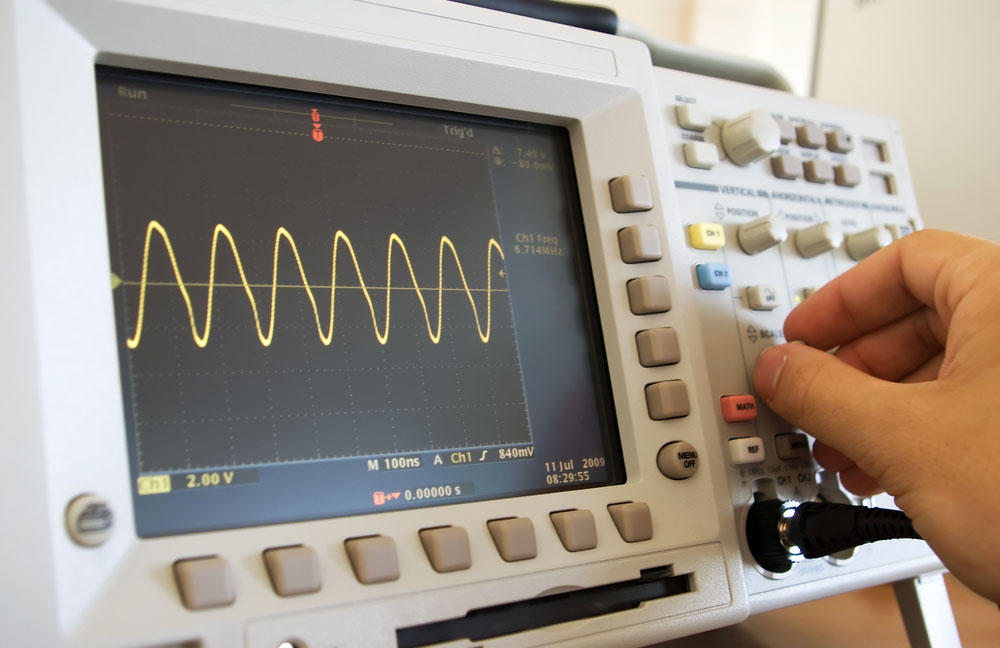}
    \caption{A typical Digital Storage Oscilloscope (DSO). Image credits: \cite{oscilloscope}}
    \label{oscilloscope}
\end{figure}

These days, the most direct way of visualizing a sound signal, is by converting it to an electrical signal or voltage. This is typically done through a microphone which has elements inside it which produce an electrical signal in response to pressure variations, called piezoelectric materials. Here, the magnitude of the pressure signal at any instant in time directly corresponds to the voltage value of the signal at that time. The electrical signal - typically quantified as a voltage - can be viewed on an instrument called an oscilloscope, where the voltage - on the y-axis - is plotted against time on the x-axis. Figure \ref{oscilloscope} shows an image of a typical oscilloscope screen. 
\subsubsection{Periodic signals and spectral analysis}
\begin{figure}[htb!]
    \centering
    \includegraphics[width=10cm]{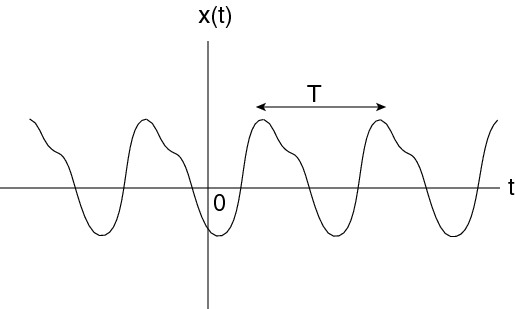}
    \caption{A periodic signal of period $T$. Image credits: \cite{periodicsignal}}
    \label{periodic}
\end{figure}

One common category of signals are ones that repeat themselves after a certain fixed amount of time. These are called periodic signals (or - more colloquially - waves), and the time after which a signal repeats is called the period. We will see that these signals play a crucial role in our understanding of sound. An example is seen in Figure \ref{periodic}, where we see a periodic signal of period given by $T$. 

\begin{figure}[htb!]
    \centering
    \includegraphics[width=10cm]{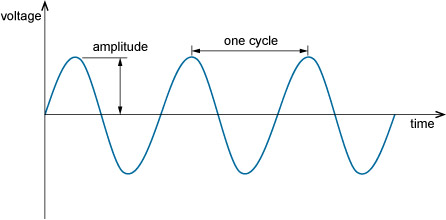}
    \caption{A sine wave signal with amplitude and period (one cycle) marked. Image credits:\cite{sinewave}}
    \label{sine}
\end{figure}

There are many common examples of periodic signals (triangle waves, square waves etc.), but one special set of such periodic signals are sine and cosine waves (Figure \ref{sine}). These sines and cosines are the same ones you might have encountered in a trigonometry class, but their exact form is not important for our purposes. 

The parameters of a sine wave are its amplitude and period  - all marked in Figure \ref{sine}. The period can also be written as a frequency, which is the inverse of the period. The unit of frequency is Hz or s$^{-1}$, which represents how many periods can fit in one second of the signal.
Our ears are especially good at distinguishing these sine waves from other signals. In fact, we hear sine waves as a pure note, with the pitch of the note determined by the frequency of the note. For example, if 'Sa' has a particular frequency, 'Ri' has a higher frequency, and so on, until the 'Sa' of the next octave which has exactly twice the frequency of the initial 'Sa'. Tuning forks, for example, are instruments which produce pure sine waves of a given frequency.

\begin{figure}
    \centering
    \includegraphics[width=10cm]{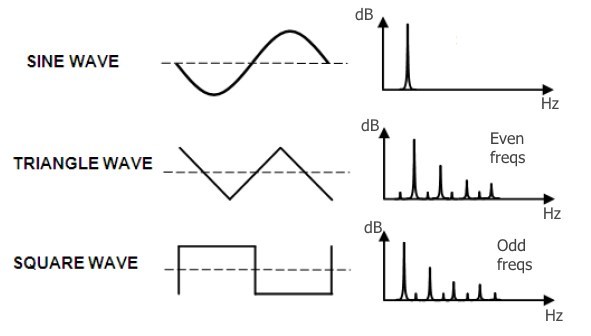}
    \caption{Various periodic signals on the left and their frequency spectrum sketches on the right. Image credits: \cite{periodicspectrum}}
    \label{periodicspectrum}
\end{figure}

The remarkable quality of sine/cosine waves, that can be proved mathematically, is that any periodic signal can always be broken down into a weighted sum of sines and cosines of different frequencies (i.e) sines and cosines of various frequencies added up in different proportions known as weights. This process is called a Fourier Transform, and is a common tool in science and engineering. Therefore, instead of representing a signal as a voltage versus time, we can instead represent it as the weights (in voltage) of the sine waves that make up the signal (on the y-axis), and their corresponding frequencies (on the x-axis). This representation is typically called the spectrum of a signal. It is shown for sine wave, triangle wave and square wave signals in Figure \ref{periodicspectrum} - the graphs on the right are the spectra of the signals on the left. We see, for example, that a sine wave has a single peak corresponding to the frequency of the wave.
Spectra for a few common periodic signals is compared to that produced by a stroke on the mridangam in Figure 6.5 in Musical Excellence.

\subsubsection{What does this have to do with music?}
\begin{figure}[htb!]
    \centering
    \includegraphics[width=10cm]{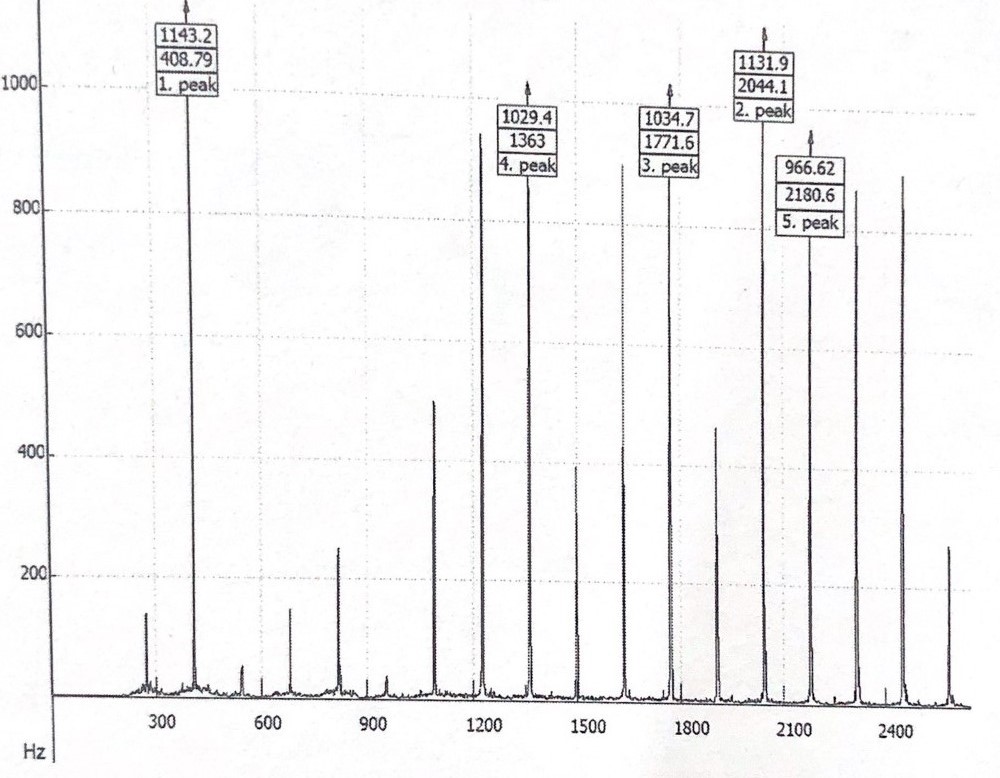}
    \includegraphics[width=10cm]{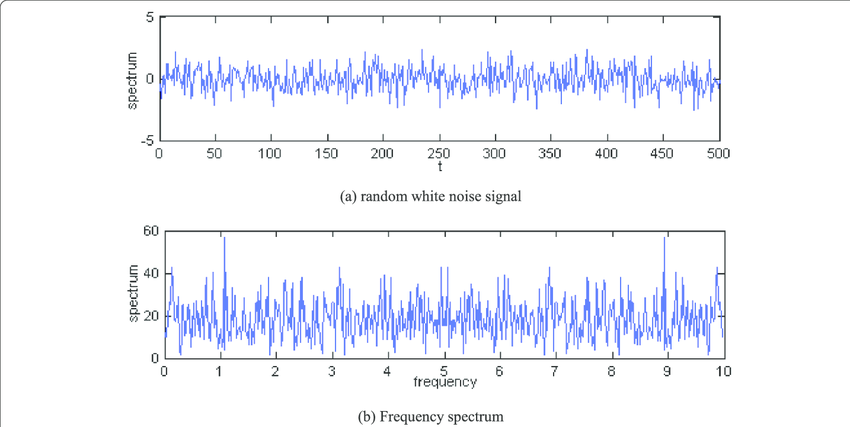}
    \caption{Top: FFT spectrum of a note on the tampura (taken from Musical Excellence). Bottom: White noise signal and corresponding FFT spectrum. Image credits: \cite{Zhang2016}}
    \label{spectcomp}
\end{figure}

\begin{figure}[htb!]
    \centering
    \includegraphics[width=10cm]{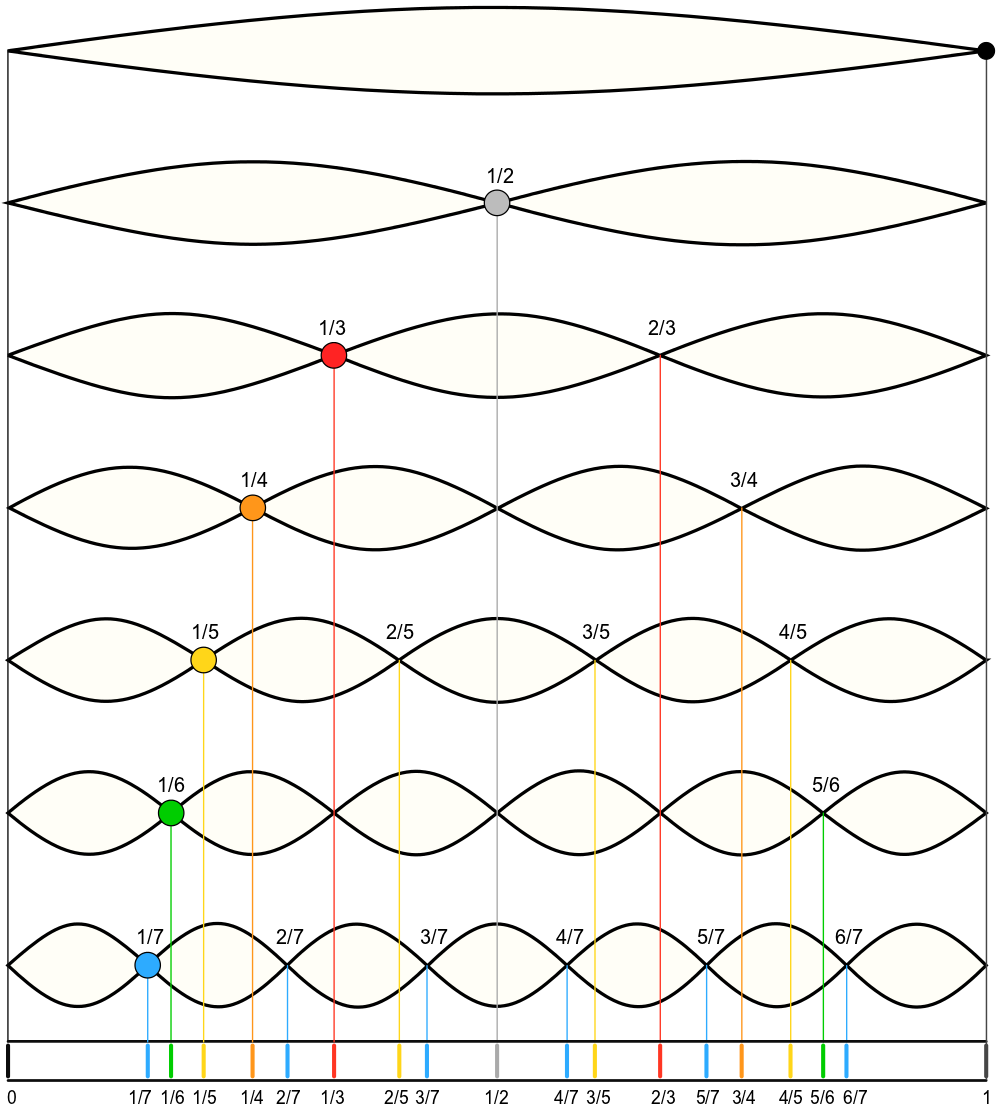}
    \caption{A series of harmonic sine waves - from the top, the fundamental or first harmonic, second harmonic or first overtone, and so on until the bottom. Image credits: \cite{harmonicseries}}
    \label{harmonicseries}
\end{figure}

\begin{figure}[htb!]
    \centering
    \includegraphics[width=6cm]{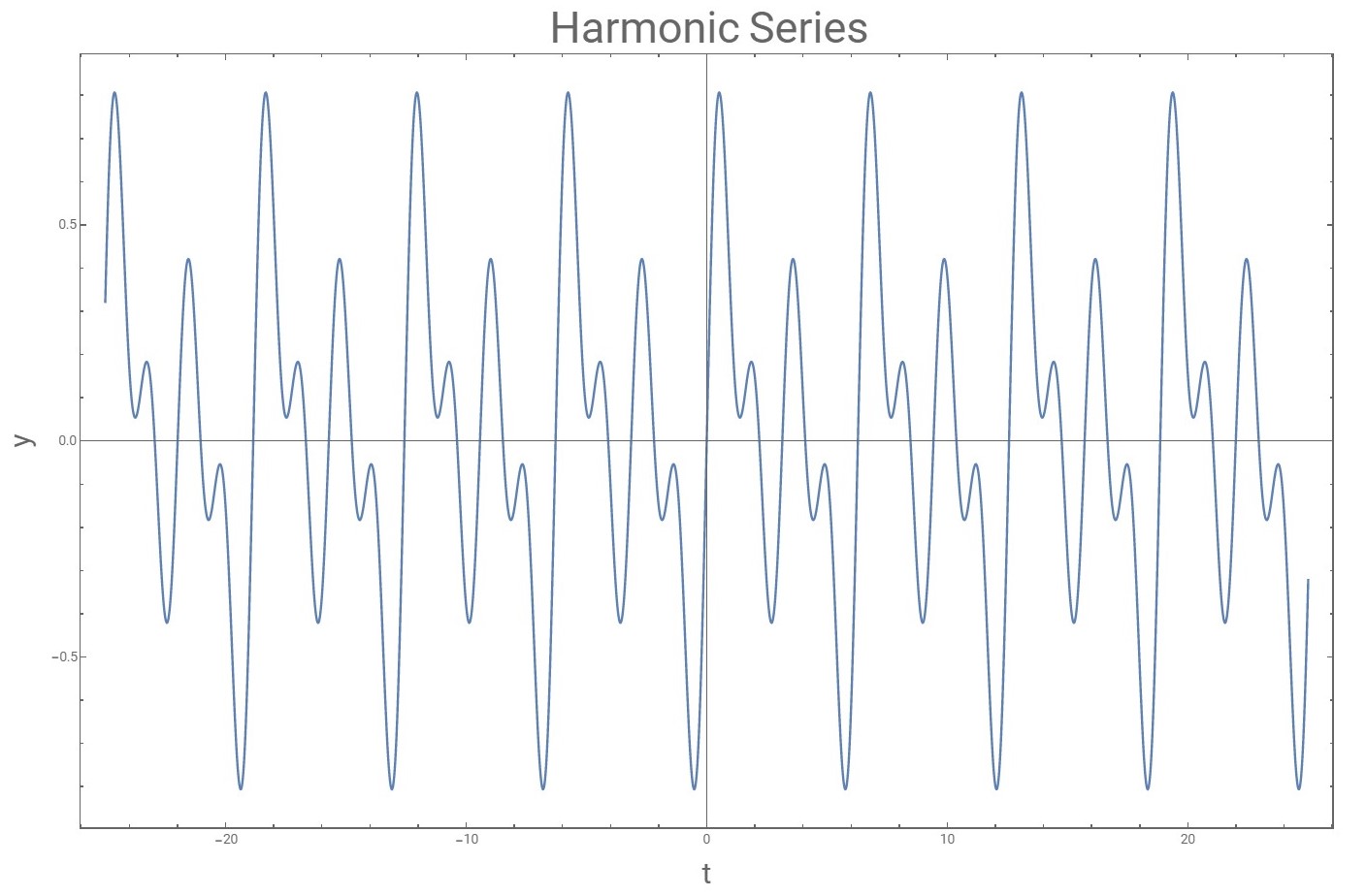}
    \includegraphics[width=6cm]{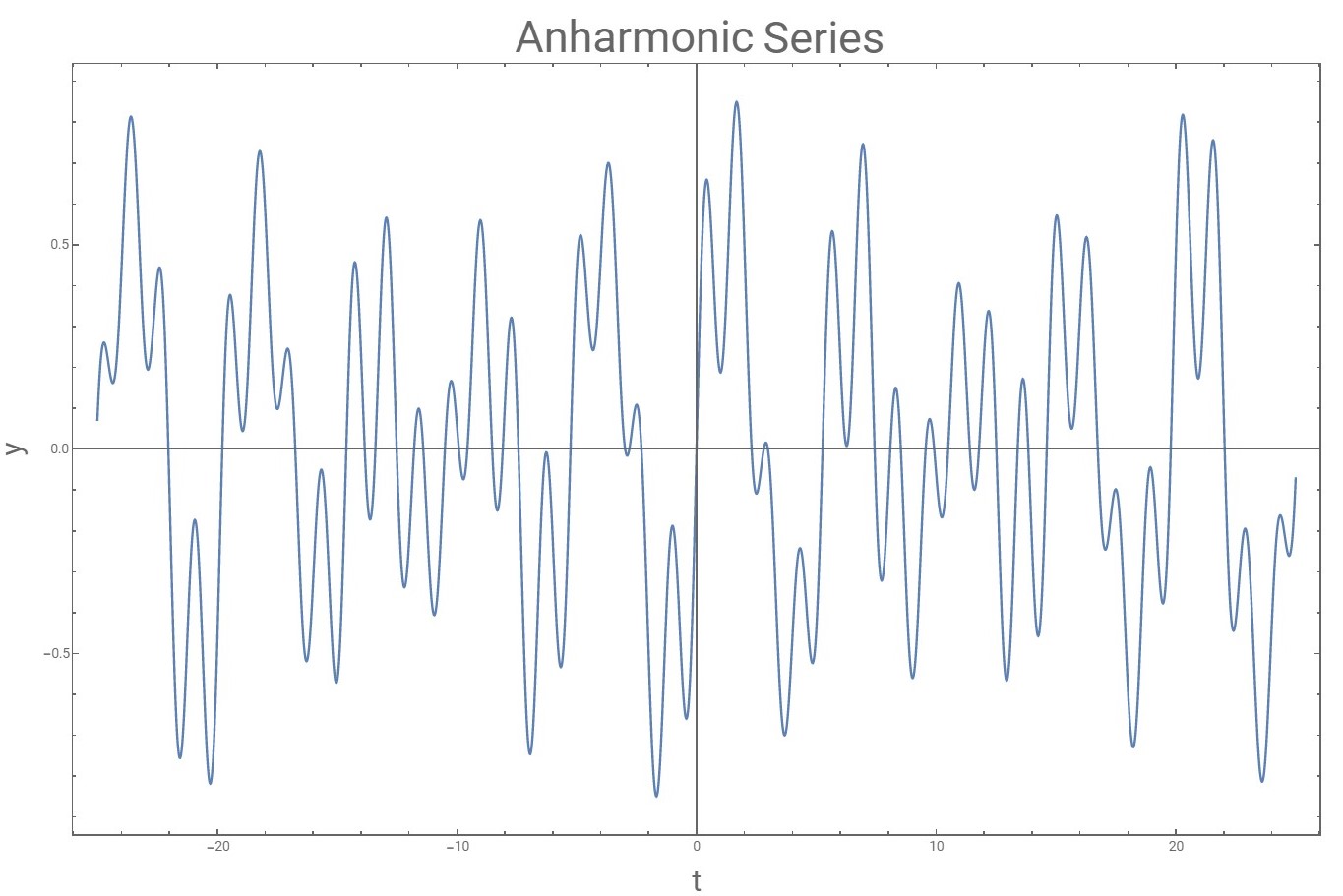}
    \caption{Left: Resultant wave of four sine waves with frequencies in harmonic ratios (1:2:3:4) in arbitrary amplitude ratios. Right: Resultant wave of four sine waves with frequencies in an arbitrary anharmonic frequency ratio (1:1.27:3.82:4.27) and the same amplitude ratio as the harmonic case \cite{combiningwaves}}
    \label{harmonicwave}
\end{figure}

\begin{figure}[htb!]
    \centering
    \includegraphics[width=10cm]{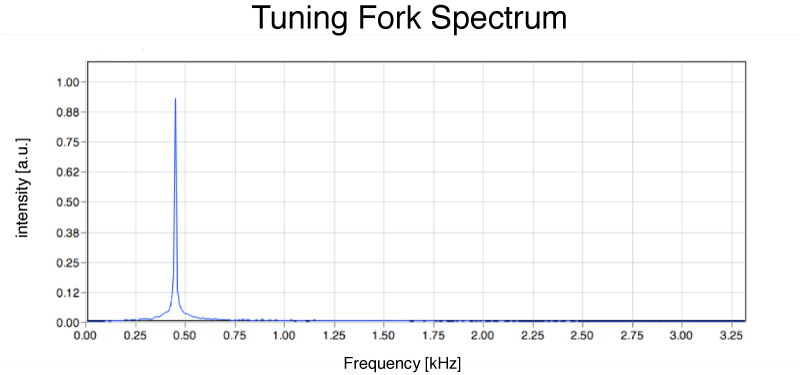}
    \caption{FFT spectrum of a note on a tuning fork. Image credits: \cite{tuningfork}}
    \label{tuningforkspectrum}
\end{figure}

As mentioned before, our ears have a keen ability to distinguish music from noise. Therefore, in exploring how this is possible, it is reasonable to expect that 'music' and 'noise' have distinct spectral signatures. In Figure \ref{spectcomp}, we compare the spectrum of a musical note generated from a tampura (taken from Figure 5.5 in Musical Excellence) and noise generated from a white noise machine. We immediately observe a few points of difference. The musical note is only composed of sine waves of a few frequencies, which are called the harmonics. Noise, on the other hand, is composed of sine waves of many frequencies and one cannot distinguish distinct harmonics. 
If we look at the spectrum of the tampura note in Figure \ref{spectcomp}(a) closer, we also observe that all subsequent harmonics have frequencies which are integer multiples of the first. For this reason, the first harmonic is called the fundamental, and higher harmonics are called overtones. This is an important feature of musical or harmonic tones. These harmonic sine waves have similar nodes (points where the amplitude of the signal is zero - seen in Figure \ref{harmonicseries}), and hence sound similar to our ears. When these harmonic series are combined in various ratios, the resulting wave retains the essential features of a sine wave and hence sounds pleasing to our ears (see Figure \ref{harmonicwave} and caption for details). Whereas, for an anharmonic sequence of peaks, the combined wave looks more similar to the wave for noise as it lacks the periodic features of the former and therefore, our brain interprets it as such. Another feature of harmonic peaks is that, due to the aforementioned structure in the nodes, they interfere constructively with each other and are much more stable than anharmonic peaks. Thus, they are able to 'sustain' for much longer.   
It is also a widely accepted fact that the number of overtones contribute to the texture of the musical note, and in general, the more the number of observable overtones, the more complex and pleasing the texture of the musical note. In musical parlance, these sine waves with frequencies in integer ratios correspond to the same note in different octaves. Hence, we see that a note played in a particular octave has overtones in higher octaves. As an illustrative example of this, we see in Figure \ref{tuningforkspectrum}, a note ('Sa' for example) played on a tuning fork used in laboratories as opposed to the tampura in Figure \ref{spectcomp}. As we see, 'Sa' by itself only corresponds to a sine-wave of a particular frequency. However, to obtain the pleasantness of a 'Sa' played on a tampura, we also need to produce overtones of the sine-wave corresponding to 'Sa', but in higher octaves, at specific amplitude ratios. The combinations of the fundamental and overtones, at these specific amplitude ratios (which, of course, vary according to the instrument used and even the musician), produced by the same instrument, combining together give us the pleasant 'Sa' played on a tampura. Possibly, it is this texture, provided by the higher harmonics of the particular note produced in ratios unique to the instrument that produces them, that provokes in us the subjective experience of 'naadham'.

\subsubsection{Producing a musical note}
Given that it is desirable to produce musical notes with a maximum number of harmonics, how do we achieve this? For this, we have to understand how sound is produced from instruments. Sound waves are usually produced by vibrating membranes, with the shape of the sound wave being defined by the shape of the membrane after exciting it. Thus, our objective is to look for systems which can produce multiple harmonics upon exciting them in specific ways. 

\begin{figure}[htb!]
    \centering
    \includegraphics[width=10cm]{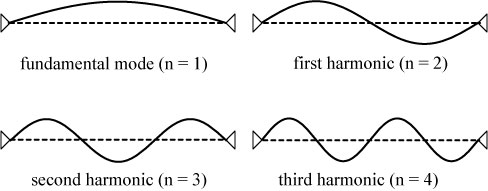}
    \caption{First four modes of a string clamped on both ends. Image credits:\cite{stringmodes}}
    \label{stringmodes}
\end{figure}

To solve this seemingly difficult problem, we look at the modes of vibration of an instrument. These are the only possible sine-wave deformations that can be done to the instrument that are stable, each characterized by the amplitude and wavelength of the sine wave. In this case, wavelength replaces period since there is a space axis instead of time, but the wavelengths are directly related to the period (and hence the frequency) of the sound signal produced. Some of these modes for a string clamped at both ends (similar to a guitar or violin) are shown in Figure \ref{stringmodes}. An interesting result, whose intuition can be understood based on our earlier discussion on the properties of sine waves, is that any possible stable deformation of an instrument is a combination of these modes of vibration. Therefore, the typical spectrum of the sound produced by such an instrument will have peaks at the frequencies corresponding to the sound produced by the modes of the instrument. If these frequencies - termed as natural frequencies of the instrument - are in a harmonic series (i.e) they are in integer multiples of the lowest frequency (the fundamental), the note that we hear upon exciting the instrument will be musical, with the texture of the note depending upon the number of harmonics sufficiently excited.

\subsubsection{What of drum instruments?}
You will notice that so far, we have not at all discussed one important aspect of music typically associated with drum instruments like the mridangam - rhythm. Drum instruments typically involve a two-dimensional membrane stretched across a surface, which is excited by striking it either using the hand or an implement like a drum stick. These two-dimensional membranes can be struck at varying pace, upto several times a second, and and their primary role is rhythm and percussion, which involves a separate discussion on the time evolution of musical and non-musical notes. However, their ability to produce musical notes is typically very limited. This is because, unlike the one-dimensional string, the uniform two-dimensional membrane does not produce modes whose frequencies are in a harmonic series. Typical ratios of the natural frequencies of the modes of a circular drum, as seen in Western percussion instruments, are in the ratio 1:1.59:2.14:2.30:2.65:3.16:3.50 etc. 

One unlikely advantage of this fact is that, because of the lack of resonance or constructive interference between modes which is seen in the harmonic case, the notes produced by a drum do not sustain for a long period of time and decay fairly quickly. This allows the human ear to distinguish upto several notes played within a second as independent notes, which is not possible if the notes sustain for a longer period since the next note will start before the first note decays and one hears a combination of both notes. This fact particularly helps the drum players to play complicated rhythmic patterns and fast tempos, which are still discernable to the listeners. An image of the first few modes of a circular membrane are shown in Figure 5.6 of Musical Excellence, and are also reproduced later in this review, in Figure \ref{kucchithool} (These are modes for the case of a mridangam, which we will see is not the same as a uniform circular membrane. However, the shape of the modes remains similar).

\subsection{A unicorn in the world of percussion instruments}
As was explained in the previous section, it is not possible for a uniform two-dimensional membrane to produce harmonics since its mode frequencies are not in a harmonic series. But what would happen if we make the circular membrane non-uniform? Could we engineer it to produce harmonics? This was the ingenious thought that was had by makers of Indian percussion instrument makers thousands of years ago, without the benefit of modern technology and understanding. They found that, by adding an additional mass around the centre of a uniform circular membrane (or 'centrally loading'), they were able to produce notes which sounded harmonic or musical. Indeed, we will see in this review that centrally loading a uniform circular membrane in particular ways does make the anharmonic modes of a uniform membrane into a harmonic series. Thus, in the world of percussion instruments, the mridangam and tabla are unique because they are drum instruments used for percussion, which are also able to generate musical notes with many harmonics.

Given this uniqueness in the method of producing musical notes, one also expects the musicality of these notes to be similarly unique. We will take the case of the mridangam and explore the unique melodic character of the notes produced, and the various factors involved to produce a tone of high quality.
\begin{figure}[htb!]
    \centering
    \includegraphics[width=5cm]{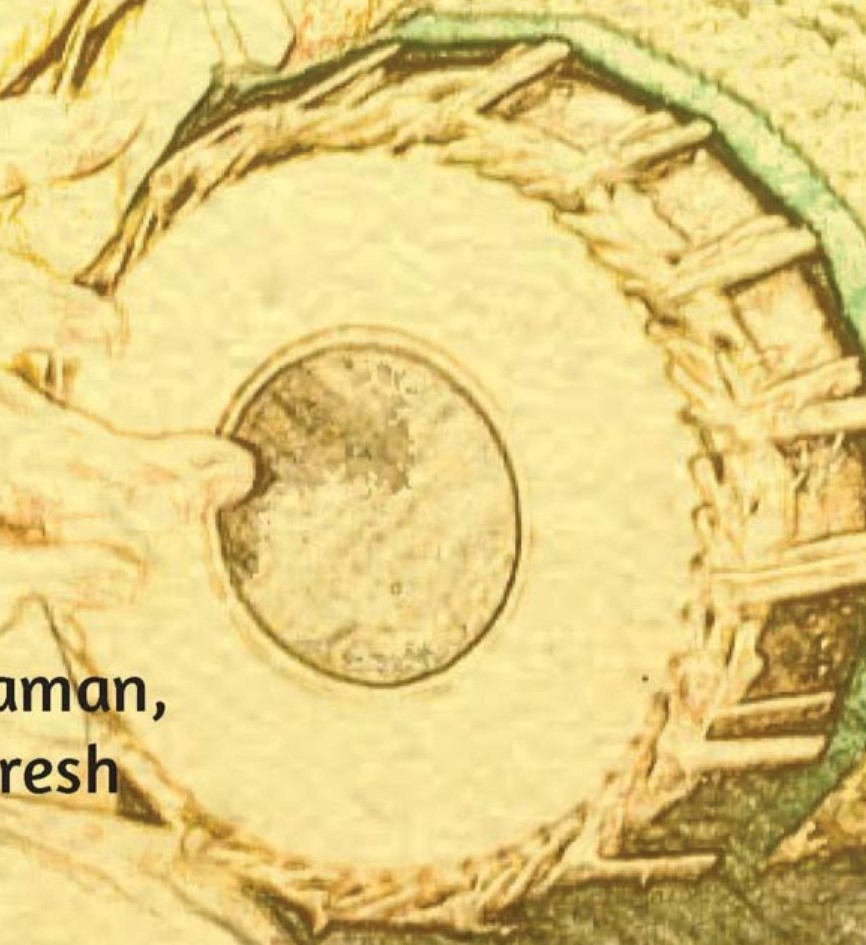}
    \caption{Image of right membrane of the mridangam taken from the cover of Musical Excellence of Mridangam by Dr. Umayalpuram K Sivaraman, Dr. T Ramasami and Dr. M.D. Naresh , with the prominent black patch seen which is responsible for central loading}
    \label{fig:my_label}
\end{figure}

\section{Can Science and Aesthetics Coexist? - History of Scientific Work on the Mridangam}
In the general public's notion, art and science have always been seen like oil and water - opposing forces that do not mix. This is mainly because art deals with a subjective experience of aesthetic beauty and science deals with objective empirical observations that are (as far as possible) independent of the subjectivity of the scientist. While this fact will always keep a separation between these disciplines, they are also - in a way - united by their objectives. Both art and science are motivated by a sense of wonder and exploration that seems to be innate to humans. Science pursues this wonder through exploration of the beauty and mysteries of the universe around us, while art pursues this through the exploration of the internal human experience of aesthetic beauty in itself. With this perspective, it is maybe not a surprise that many great scientists loved to spend time with their musical instruments, and many great artists have taken an interest in science.

Musical Excellence of Mridangam (Musical Excellence for short) is a treatise born out of such a shared sense of wonder and exploration between eminent scientists Dr. Ramasami and Dr. Naresh, and the mridangam maestro Padma Vibhushan Dr. Umayalpuram K. Sivaraman. Given that mridangam is an ancient art and the science behind how the instrument works is still relatively recent, Musical Excellence - throughout the treatise - uses inputs from Dr. UKS as a starting point and initial hypothesis for all of its scientific investigation, and builds on that though rigorous scientific analysis, to make new discoveries and give new insight into the workings of this instrument and the creation of 'naadham'. For this unique research setup to work, it requires an artist who has a very good grasp of the science behind the instrument, scientists who have a keen ear for music, and open conversation and flow of ideas between the two.

The first chapter of Musical Excellence thus aptly begins with conversations with both Dr. UKS and Dr. Ramasami which help to elucidate the objective and motivation behind the treatise and also their philosophy of mridangam, research, music, art, science and many more, with many interesting anecdotes along the way. The conversation with Dr. Ramasami establishes the technical base of the treatise, where he first explains the scientific background for research into the mridangam, and goes on to expound on the various scientific tools used in the current research work and summarizes the main findings of the project. Along the way, we also get to learn his perspective on naadham and the relationship between science and music. The conversation with Dr. UKS explores the concept of musical excellence in the context of the mridangam and includes discussions on his philosophy of naadham and mridangam, the role of the mridangam in a concert, his role as a guru and many more practical issues like sustaining energy for a long concert, various types of mridangams and methodologies of accompaniment. Amongst these are sprinkled various illuminating anecdotes from doyens of Carnatic music like Sri Semmangudi Srinivasa Iyer, Sri Arriyakudi Ramanuja Iyengar, Sri Pallakadu Mani Iyer, Smt. D.K.Pattamal and many more. In addition to this, Dr. UKS also explains his motivation behind his ongoing pursuit of scientific research into the mridangam and the innovations that he has made thus far, including the fibre-glass mridangam, before going on to outline his objectives for the current research project. 

In this chapter, we will look at the existing body of literature on the science of mridangam, sparked by the pioneering work of Dr. CV Raman, which will lead us to understand the level of scientific understanding that currently exists about this instrument, and what questions still need to be answered.

\subsection{Laying the Foundations - Dr. C.V Raman}
Dr. CV Raman - the scientist of great repute for his discoveries on waves both light and sound - through his keen musical intuition realized that the Mridangam was not only a percussion instrument but a melodic one as well. This motivated his foray into the harmonics of Mridangam strokes\cite{Raman1920,Raman1922}, and culminated in a landmark paper in 1935 \cite{Raman1935}, where he explains the results of his experiments performed 1919 onward. Raman was the first to notice that the mridangam could produce upto five harmonics, thus beating the no-go theoretical result about the inharmonicity of 2D membranes. Raman was further able to demonstrate the different shapes of the membranes corresponding to each of the modes (Figure 3.1 in Musical Excellence ) and recognize the role of a skilled player in exciting some of these modes. Additionally, he hypothesized the role of central loading (which was later proven by theoretical works) and the heavy wooden shell in the generation and sustenance of these harmonic modes, and to top it all off, also noticed that the fundamental mode is shifted in frequency by about 10 percent (The actual shift is by seven percent). It is amazing that Raman was able to find out so much without the aid of any electronic oscillators or spectral analysers and only using his keen ears and physical intuition. The shifted fundamental is a curious fact that has attracted the interest of future theoretical investigation. It also means that the mridangam is tuned using the second harmonic mode in the series, since the first harmonic's frequency is offset. 
\subsection{Building on Raman's work}
Following Raman's breakthrough findings on the harmonic nature of notes produced by the mridangam, theoretical efforts by Ramakrishnan et.al in 1954-57 \cite{Ramakrishna1954} modelled the black patch as a sharp increase in the density of the membrane below a certain radius. With this model they were able to reproduce the harmonic series of modes and also the degeneracy in modes observed by Raman, and further predict values for sixth and seventh overtones that were not observed by Raman. In 1994, Siddharthan et.al \cite{Siddharthan1994} were able to calculate how exactly the central region has to be loaded in order to produce the harmonics observed by Raman. As part of their study, they also observed that the fundamental was shifted by 7 per cent and the rest of the series was harmonic. Additionally, they found that the dheem stroke (see section 5.1.1 of this review) primarily excites this shifted fundamental, whereas the chappu stroke typically used to tune the mridangam to the pitch of the main artiste primarily excites the second harmonic. This would imply that the mridangam is tuned using the second and higher harmonics, but our ears are able to perceive the fundamental that would correspond to this harmonic series and match the frequency to the pitch of Sa. Further, in 2000, Malu and Siddharthan \cite{malu2000} presented a model which accounts for the actual loading pattern of the black patch on the mridangam and reproduce the harmonic frequency ratios. Further studies primarily focus on comparing tonal properties between the mridangam and tabla, and these are also catalogued in Chapter 6 of Musical Excellence.
\subsection{Current Research Questions}
After Raman's initial work, there has been very few subsequent experimental works that have emerged, which would reflect the level of theoretical understanding we currently have about the origin of harmonics in the mridangam. Systematic experimentation is required to distinguish between the different theoretical models of harmonic production, and quantify the role of different structural factors like the wooden shell and membrane materials, and human factors like the expertise of the artiste, in producing tonal beauty. Apart from this, there is ongoing effort to standardize the fabrication methodologies of the mridangam, find alternative materials for constructing the mridangam, and design a travel-friendly mridangam.  It is with the goal of addressing these questions that the authors of Musical Excellence of Mridangam pursue their work, the results of which are summarized in the subsequent chapters.

\newpage 

\section{Musical Excellence of Mridangam - Materials and Making}
The fourth chapter of Musical Excellence deals with the materials, structure and construction of the mridangam, and investigates how these various factors are involved in producing the tone of the mridangam. After introducing the structure of the mridangam, Musical Excellence explains the traditional process of constructing the mridangam in minute detail. Following this, Musical Excellence delves into the six categories of materials used in the mridangam and investigates the important material properties of each. Having established these, it also investigates the elements of construction that might be important to producing musical notes in the mridangam.

\subsection{Structure of Mridangam}
The basic structure of the mridangam consists of a wooden shell with two multi-membrane assemblies on either side, which are held taut by straps connecting them. On the right hand side or valanthalai - which contains the black patch and is hence the major source of harmonics - the membrane consists of multiple layers, the first of which is a disk of buffalo skin parchment called ulkaraithattu, which is directly tied to the wooden barrel through buffalo hide straps. Upon this is placed a full membrane typically made of goat skin parchment, which also contains the black patch. This is the primary membrane whose vibrations are responsible for the harmonic notes - hence called paatuthattu or kottuthattu. Upon this membrane, annular rings made from cow skin parchment are placed to create a space between membranes which is filled with different materials to modify and fine-tune tonal quality. The final layer - which is a disk typically made out of cow skin parchment - called vettuthattu - is added. The annular space between the kottuthattu and vettuthattu is filled with different materials like pine wood sticks or pieces of black patch material to fine-tune and control the harmonics produced. The black patch itself is made of a powder of ferric oxide and is applied on the kottuthattu in sequential layers to create a density gradient that is important to create harmonic modes - as seen by Malu and Siddharthan \cite{malu2000}.
The left membrane of the mridangam, or edanthalai, is typically used for playing base notes, as opposed to the treble notes played from valanthalai. This membrane is made from buffalo parchment which is typically thicker than the right membrane. While this membrane is typically not centrally loaded with a black patch since its primary role is not the production of multiple harmonics, it is usually centrally loaded while playing through application of a paste of semolina or cream of wheat (rava) to create base notes that have increased sustenance and nevertheless produce some harmonic behavior.
\begin{figure}[htb!]
    \centering
    \includegraphics[width=10cm]{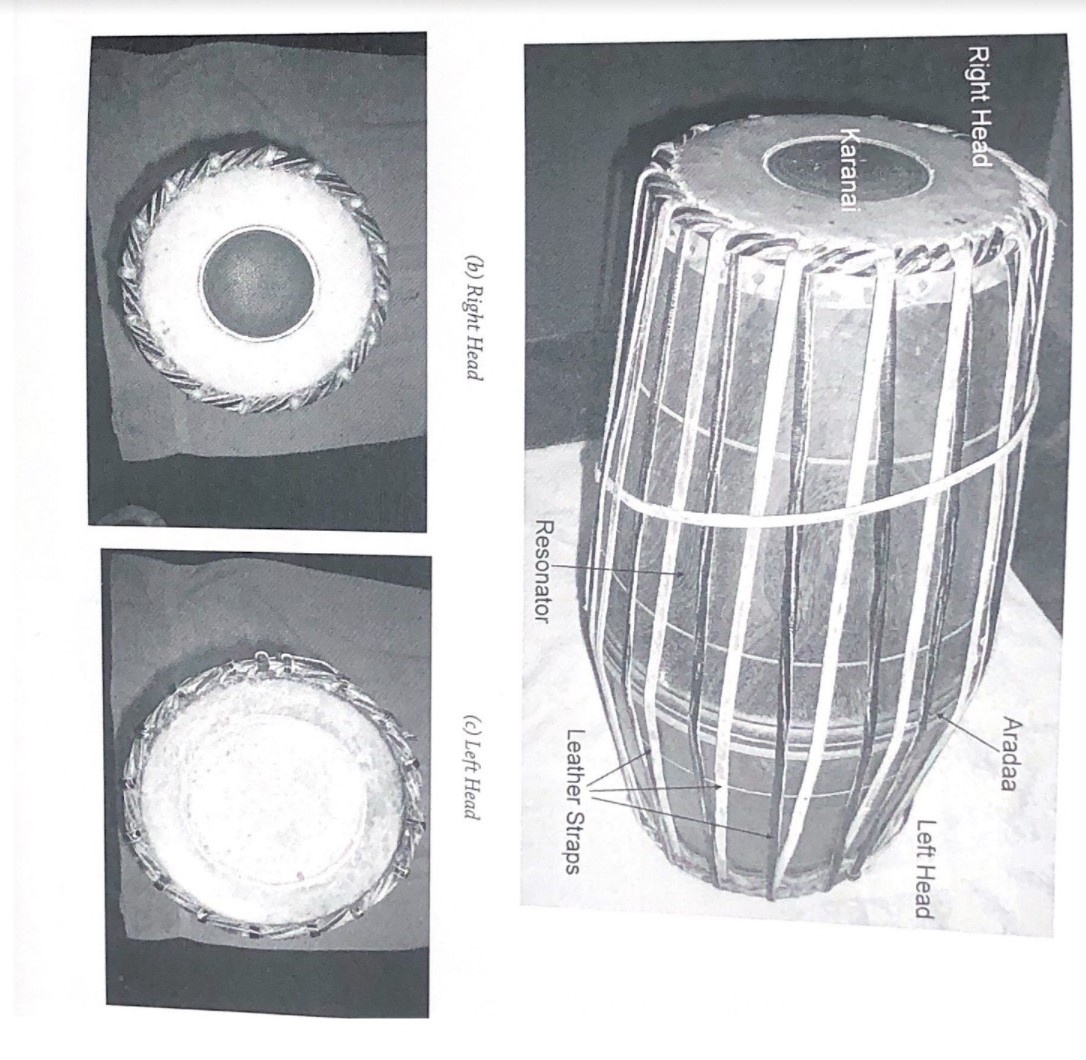}
    \caption{Structure of the mridangam. Image taken from Figure 4.1 of Musical Excellence.}
    \label{fig:my_label}
\end{figure}
\subsection{Traditional Construction of Mridangam}
The main parts of a mridangam are a central wooden barrel which serves as a resonating column - typically in three sizes -22", 23" and 24" - whose openings are covered by a multi-layered assembly of parchments made from goat, cow and buffalo skin. The process of making a mridangam starts from the trunk of a jackfruit tree, which is machined in order to make wooden barrels of required height and diameters. For the left side membrane, buffalo skin parchments are soaked overnight, and the layers are carefully assembled by tying together with hoops and thongs. The right side membrane is also assembled in a similar fashion. The buffalo hide straps are used to tie the two membranes to the wooden shell in a plaited manner, through holes made on the edges of the membranes. For the right side membrane, since equal distribution of forces is critical, additional holes are punched and a rope is run through them circumferentially - the number of points of insertion for the right membrane is 48, while it is 36 for the left membrane - in order to equally distribute the tension and allow for fine adjustments of shruthi. Once the parchments are set on the shell and dried, the buffalo hide ropes are removed. The process described thus far is called 'poi var pidithal' since the applied tension is not retained once the parchment settles on the shell. Finally, the left and right membranes are joined through buffalo hide rope which runs through 16 points along both membranes and is held taut so that the right membrane has the necessary tension. The black patch is then applied to the exposed region of the kottuthattu layer-by-layer, after cooked rice is applied as an adhesive to the membrane. The annular space between the kottuthattu and vettuthattu is then filled using radial strips of reeds (kucchi) or pieces of black patch material (thool) according to the artiste's preference.

At the beginning of the fourth chapter, Musical Excellence explains this process - which has been gleaned from extensive conversation and observation of skilled artisans like Sri Johnson - in full detail, and includes images which show the various steps being performed. Throughout the fourth chapter, the authors consider the potential steps towards standardization of this construction process and demystifying the myths that have invariably passed down to us as well, which could - for example - lead to a future where we could design and build this instrument using sustainable materials or make a transport friendly version, while preserving the unique acoustic qualities. As a first step in this direction, the authors suggest the use of Fast Fourier Transform (FFT) techniques to extract and characterize the spectra of different strokes played on the mridangam, along with time-decay profiles to quantify the sustenance of different harmonic peaks. FFT is a computational technique used for performing the Fourier Transform and extracting the spectrum of a signal (see Chapter 1 of this review for an introduction to Fourier Transforms and spectral analysis).

\subsection{Rational Choice of Materials}
There are six classes of materials in a traditional mridangam - wood (main body), parchments (membrane), minerals (black patch), adhesives (binding black patch to memberane), straps (binding membranes with desired tension) and materials for filling the annular space (kucchi vs kappi). The authors study each of these six classes of materials separately, and present rational choices for materials based on such analysis. In addition to allowing us to make rational choices about the materials currently used in a mridangam, this section establishes the scientific method to studying these materials used in the mridangam, thus establishing the grounds for discovering new such materials which satisfy certain requirements like sustainability or transport-friendliness or superior acoustic properties, and move us towards standardization. Later, In the fifth chapter, dealing with tonal properties, Musical excellence also addresses a few popular material-related issues like goat vs cow parchments and kucchi vs thool mridangams.

\subsubsection{Wood}
Wood is the preferred choice of resonating column for the mridangam, and hence represents one of the most important material choices. One of the unique properties of wood is that it has different and independent mechanical properties along three different mutually perpendicular directions (Orthotropic), due to the formation of growth rings. This is well illustrated in Figure 4.6 in Musical Excellence Musical Excellence, where we see different patterns created by the growth ring fibres along different mutually perpendicular surfaces, leading to unique mechanical properties along each direction. In the case of the mridangam, the wooden shell is made in a direction axial to the run of the fibres. To understand the orthotropic effect, the authors conduct a study of the three-dimensional propagation of sound for different types of wood typically used in the mridangam. Musical Excellence further introduces the possible design choices in wood through analogy with the role of wood in violin construction, particularly the famed Stradivari. The wooden shell is made in the form of a frustum to ensure that the buffalo hide straps are under tension, and it typically comes in lengths of 22",23" and 24" - depending on the required shruthi.

To study the role of wood, Musical Excellence uses the parameter of Sound Radiation Coefficient (SRC), which has been previously used in studying the acoustic qualities of wood used in xylophones and soundboards. SRC is a measure of the vibrational response of a material to an input pressure signal (i.e) sound wave. Higher SRC values are typically seen as ideal for musical instruments as these correspond to more loudness for a given applied force. 

Another such quantity used by the authors is Impedance (Z), which quantifies sound propagation through a material. If sound moves from one system with impedance Z1 to one with impedance Z2, the transmitted sound intensity drops significantly if the difference between Z1 and Z2 increases. Thus, for efficient sound intensity transmission, we want the impedance of different parts of our system to be as close together as possible. In this case, the authors compare the impedance of different directions of the wood (as mentioned above), for different types of wood used. 

Both these factors depend on the density,velocity and Young's Modulus (a parameter which quantifies response to external stress) of the wood, and these are subsequently measured in order to calculate SRC and Impedance.

On a more practical note, it is worth to compare the hardness of wood to ensure ease of machinability - as a step towards standardization. The authors thus study this factor through the Janka Hardness Test, which is typically used in the case of wood.

Through this study, the authors are able to conclude that jackfruit wood is the favorable choice among the different types of wood studied, due to its high SRC and similar Impedance in different directions. This seems to agree with conventional wisdom, as traditional makers have favoured jackfruit wood for the construction of the mridangam barrel. The comparatively low density of this wood also reduces the total weight of the instrument.

\subsubsection{Parchment} 
Parchment has emerged as the membrane of choice for the mridangam, and is also used for the straps which adjust the tension of the membrane (typically made from buffalo hide). In Musical Excellence, we see that parchment made from different animals, and even different part of the same animal (particularly, back and belly), can be used to generate different musical outcomes based on their unique properties. To bring these on a common footing, Musical Excellence categorizes these different skins based on the thickness of the skin, the diameter of the fibre bundles that run along the skin, and the angle at which these fibres are weaved together. The effect of processing these skins by treating with lime (calcium hydroxide) is also explored. It is noted that skin from the back region of the goat might be more suitable for producing resonances due to the better compaction of fibre bundles in that region, and hence it is used for the kottuthattu of the right membrane of the mridangam which contains the black patch and is used to create the treble notes. On the other hand, the skin from the belly region, having lesser compaction of fibre bundles, is more suitable for the left membrane which is used to produce bass tones.

The study of different parchments is done in Musical Excellence through comparison of their visco-elastic properties, which quantifies how the membranes respond to an external force (in this case, the force applied by the artiste) (i.e) whether the membrane relaxes to it's original position in the same configuration after deformation (elastic) or undergoes a permanent deformation (plastic) or returns to the same position but there is a permanent rearrangement of molecules within the membrane  (viscous). 
It is found that skin from different regions of the goat all have similar distributions of deformation across the three types discussed, which implies a significant contributions from elastic deformation which is required for a membrane material. The plastic and viscous contributions imply that the membranes deform over time and their tension will have to be re-adjusted from time to time. The buffalo straps of the mridangam allow for this to be done, in a process called 'var pidithal'.

An anisotropy is also found in the material properties of the membrane between perpendicular directions, attributed to the orientation of the fibres parallel or perpendicular to the direction of deformation. It is also found that liming marginally increases the elastic component of deformation, which might make it more suitable for sound generation. 

Additionally, Musical Excellence compares the frequency spectrum of a 'chappu' note (a method of analysis heavily used in the fifth chapter) played on membranes made from both cow and goat parchments and finds sharper resonance peaks for the goat parchment membrane, which indicates its relative suitability for resonance production and hence for its use on the kottuthattu of the right membrane of the mridangam with the black patch. This conclusion is further explored in the fifth chapter, where other strokes are compared as well.

\subsubsection{Minerals}
Minerals are used to create a black patch on the right membrane, to load it in such a way that it produces harmonic tones (as discussed earlier). Primarily, Ferric oxide sourced from rocks on river belts like the Ganges and Cauvery are used to make the black patch. Upon analyzing the composition of minerals sourced from these locations, the authors find the samples from Cauvery to be more rich in iron, whereas the samples from the Ganges are found to be more rich in Silica. The tonal consequences of these are discussed in the fifth chapter of Musical Excellence, and will be alluded to later in this review. 

Additionally, the authors also use X-Ray diffraction - a technique used to analyze the crystalline nature of a sample and find the crystal lattice structure - to analyze these samples. In X-Ray diffraction, light is shined along different directions relative to a sample, which if it is crystalline, causes the light to diffract due to its periodic structure. This creates a pattern of dots on a screen, which correspond to peaks of light intensity. Sharper peaks indicate a more crystalline nature of the sample, and the position of the peaks indicates the crystal structure. The authors in Musical Excellence compare the mineral samples taken from Ganges and Cauvery and compare it to pure Ferric oxide. It is found that the samples of black patch are indeed primarily composed of Ferric oxide, but they are not as well-ordered when compared to the pure crystalline Ferric oxide samples. It is also interesting to see the similarities and differences in the diffraction patterns, and this analysis in Musical Excellence introduces X-Ray diffraction as a method of analysing minerals used for the black patch.

\subsubsection{Adhesives}
Cooked rice is generally the preferred adhesive used for binding the particles of black patch together and the black patch itself to the goat-based parchment. With the objective of finding other alternatives to this, Musical Excellence examines various adhesive materials using standardized tests for evaluating binding properties. The generated alternatives are examined in subsequent chapters.
\subsubsection{Straps}
Buffalo hide is usually used for the straps which connect the membrane to the wooden barrel, and are used to adjust the tension of the membrane which changes the frequency of the harmonics (i.e 'shruthi' or tone of the instrument) and thus tune the instrument. These are used for their high tensile strength and durability. Musical Excellence quantifies the tensile strength of these straps and discusses the possible role of traditional processing steps like treating with curd.

\subsubsection{Annular Space} 
Annular spaces between parchment membranes on the right side of the mridangam are typically filled with insertion materials to fine-tune the harmonics produced. These are also known to either dampen or amplify different strokes played on the mridangam. These include thool (particles left behind after sieving the black patch minerals) placed uniformly throughout the annular space, kucchi (strips of palm stems) placed radially at equidistant points along the membrane, grains of sand etc. Musical Excellence quantifies the properties of the traditionally used annular materials, and discusses the musical outcomes of thool vs kucchi - a popular issue - in the fifth chapter, which will be explored later in this review. 

\subsection{Elements of Construction}
\subsubsection{Preparation of parchment}
Parchments used for the right and left membranes are prepared using standard means by soaking in water to clean and remove salts used for preservation, and treatment with lime. This is followed by removal of hair and flesh, and the added lime using ammonium chloride as an acidic salt. They are then adjusted to the required thickness and dried. Following this, they are cut into the required shape, according to their usage.
\subsubsection{Preparation of Minerals for Application in Black Patch preparation}
The black patch is one of the most important and unique parts of the mridangam. Therefore, the scientific study of its construction is of paramount importance. We saw that the introduction of a central loading (like the black patch in the case of the mridangam) to a uniform 2D membrane shifts the anharmonic frequency spectrum to harmonic. Most of the mathematical models which prove this effect have considered layers of black patch applied in a gradual manner, with sequentially decreasing diameters for subsequent layers. This results in a type of loading where the maximum load is at the centre, and it gradually tapers off along the radius as we get closer to the edge of the black patch. However, this is not how the black patch is traditionally constructed in the mridangam. Instead, layers with smaller and larger diameter are alternated in a specific way, before applying the last few layers in a sequential fashion (see Figure 4.15 in Musical Excellence for an illustration).

Musical Excellence has done a systematic study of this method of loading, through an interesting experiment with a tuning fork - which is an instrument used to produce a tone of a given frequency. In this experiment, they apply the layers of the black patch with varying densities in the same sequence as is seen in the mridangam, and measure the change in frequency of the tuning fork. It is found that, while the frequency decreases with loading as expected, there is a fluctuation in the rate of decrease. This fact has been used intelligently by artisans to produce overtones (which are initially anharmonic) that are harmonic.

Similar studies are conducted which now measure the change in frequency of different notes like dheem, nam (meetu) and chappu upon applying successive layers of black patch, and a similar decrease in frequency is found. Interestingly, the ratio of frequency between dheem and chappu is found to oscillate as more layers of black patch are applied, until it stabilizes at a value of 0.535 after a certain number of layers. This stabilization might be used to infer that sufficient layers of black patch have been applied. Also, the value of 0.535 in the ratio of frequencies of dheem to chappu (first and second harmonics respectively) is characteristic of the mridangam, due to the first harmonic or fundamental (i.e) dheem being shifted - as explained earlier (Other wise one would expect a ratio of 0.5 between these harmonics). More detailed study reveals that this ratio is accurate to within 0.05 over a range of different mridangams. Similar ratios for non-observable fundamental to dheem (discussed earlier) - 1.07 $\mp$ 0.01 and nam to chappu - 1.5 $\mp$ 0.012 are reported in Chapter 6 of Musical Excellence. This fact provides us with a method to characterize mridangams made from alternate materials and design. This would be the first test for these new instruments to pass to be considered equivalent to a traditional mridangam.

\subsection{Chapter summary}
This chapter of Musical Excellence thus examines the role of various materials used in the mridangam, and characterizes each of these materials according to their role, using scientific tools, in order to understand the relevant material properties and come up with alternatives for standardization. Additionally, this chapter also explores the various construction processes like preparation of the black patch, which are crucial to producing the unique tone of the mridangam. The next chapter further explores this unique tonal quality through similar scientific tools like Fourier Analysis.

\newpage
\section{Musical Excellence of Mridangam - Understanding the Unique Tonal Properties}
As Musical Excellence states beautifully in the first line of Chapter 5, 'the uniqueness of the Mridangam emanates from its potential to combine the generation of harmonics like string instruments, create frequency diversity like wind instruments, and offer rhythmic beauty of percussion.' Just like the seven notes of within an octave, there are seven notes played on a mridangam - each with its own unique tonal signature. The origin of the tonal beauty of these strokes is explored in Chapter 5 on Musical Excellence.
\subsection{FFT Spectrum and Decay Profile Analysis}
In order to decode the different strokes played on the mridangam and further study the effect of materials and construction, Musical Excellence uses Fast Fourier Transforms (FFT) - explained in the first chapter - to extract the spectrum of each note and make comparisons. Additionally, the authors also measure time decay profiles, which measure how the intensity of the primary harmonic peak for a particular stroke changes with time, to quantify the sustenance of the particular note. This is a very intuitive quantity to measure, and is something both artists and scientists can relate to.
\subsection{The Seven Strokes}
In order to properly characterize the tonal quality of the different strokes - given the various factors involved including background noise and expertise of the artiste - Musical Excellence uses sound recordings of these notes played individually by Dr. UKS in a studio for minimization of background noise. We see the beautiful unique signatures of the different notes in Figure 5.1 in Musical Excellence, with the 'open' notes played on the right side eliciting maximum harmonic features of sharp peaks separated in frequency by integer ratios, and the 'closed' notes on the right side showing broader features characteristic of anharmonicity. Interestingly, the notes played on the left side ('thoppi') also show harmonic features, though these are broader than those of the right side. These harmonic features are to be expected since the membrane on the left side is also centrally loaded with a paste of 'rava' (cream of wheat) which can be removed after a session. These figures are also reproduced below.
 
\textbf{Dheem} \\
Dheem is the stroke played on the right side membrane that primarily excites the fundamental mode of said membrane. Interestingly, Musical Excellence finds that the frequency of this fundamental mode is shifted by 7 percent compared to the expected fundamental frequency from the rest of the harmonic sequence, and hence, when the mridangam is tuned to said harmonic sequence using higher harmonics, the tone of dheem corresponds to the Rishabham note. 
\begin{figure}[htb!]
    \centering
    \includegraphics[width=10cm]{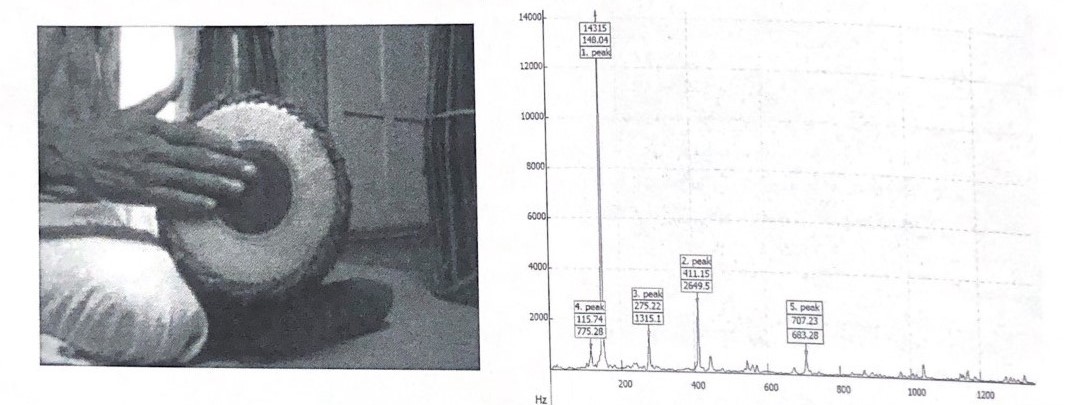}
    \caption{Dheem stroke with FFT spectrum}
    \label{fig:my_label}
\end{figure}

\textbf{Chappu} \\
Chappu (along with the Araichappu) is the primary tonal note of the mridangam, played on the right side membrane. It primarily excites the second harmonic but also has a rich polytonic structure. This stroke is used to tune the mridangam so that the expected fundamental frequency corresponds to Shadgamam or Sa of the main artiste. It is interesting that our ears are able to perceive this 'ghost' fundamental even though we are exciting only second and higher harmonics using the chappu note. The second harmonic excited using chappu sustains for the longest in comparison to other strokes (with the exception of dheem) , for even upto a few seconds.
\begin{figure}[htb!]
    \centering
    \includegraphics[width=10cm]{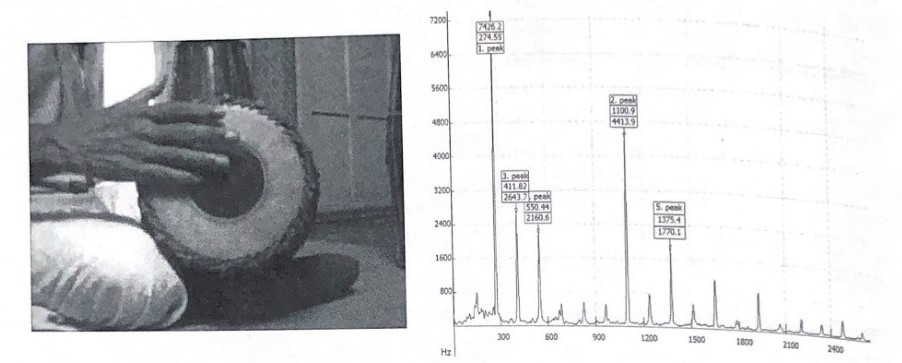}
    \caption{Chappu stroke with FFT spectrum}
    \label{fig:my_label}
\end{figure}

\textbf{Nam} \\
Nam is a stroke that is played on the periphery of the right side membrane, that primarily excites the third harmonic. It produces a metallic sound that does not sustain for very long, but is nevertheless harmonic. As seen in the harmonic peaks for this stroke which tend to be broader, it requires a skilled artiste to extract the naadham of nam.

\begin{figure}[htb!]
    \centering
    \includegraphics[width=10cm]{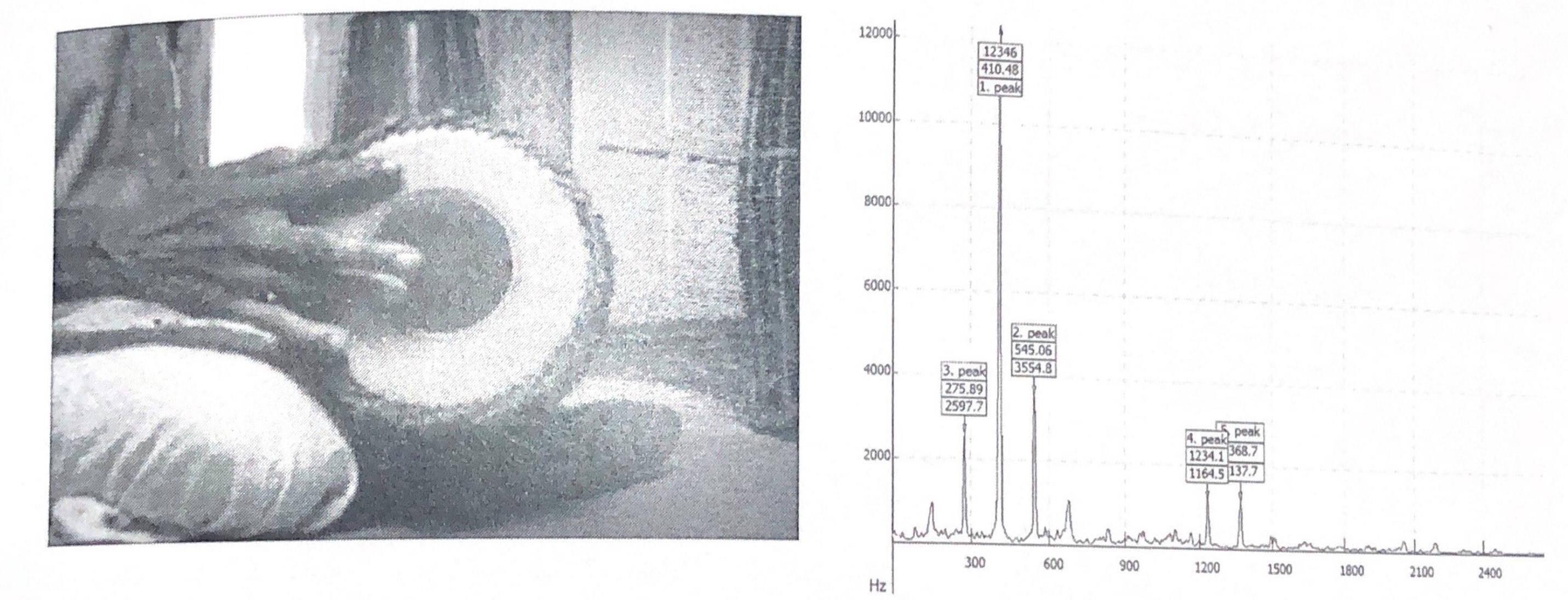}
    \caption{Nam stroke with FFT spectrum}
    \label{fig:my_label}
\end{figure}

\textbf{Araichappu} \\
The Araichappu is a stroke that is unique to the mridangam, that has no counterpart even in the tabla (as seen in Table 6.3 of Musical Excellence and corresponding discussion). The araichappu primarily excites the third harmonic, and hence corresponds to the upper octave Sa note. Additionally, while the chappu note has a soft timbre, the araichappu has a sharp, metallic timbre - even though it retains the rich polytonic structure of chappu. 
\begin{figure}[htb!]
    \centering
    \includegraphics[width=10cm]{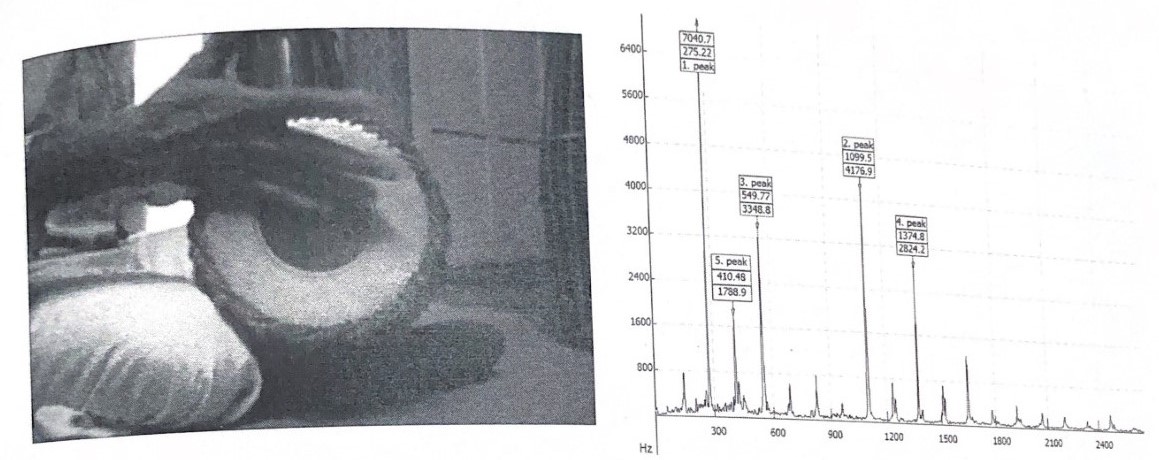}
    \caption{Araichappu stroke with FFT spectrum}
    \label{fig:my_label}
\end{figure}

\textbf{Dhi} \\
While Dhi is a closed note played on the right membrane that does not sustain for long, we see from the spectrum that it has clear harmonic peaks and is thus a tonal note. The naadham of Dhi thus requires a skilled hand to produce these harmonics even in a closed note.
\begin{figure}[htb!]
    \centering
    \includegraphics[width=10cm]{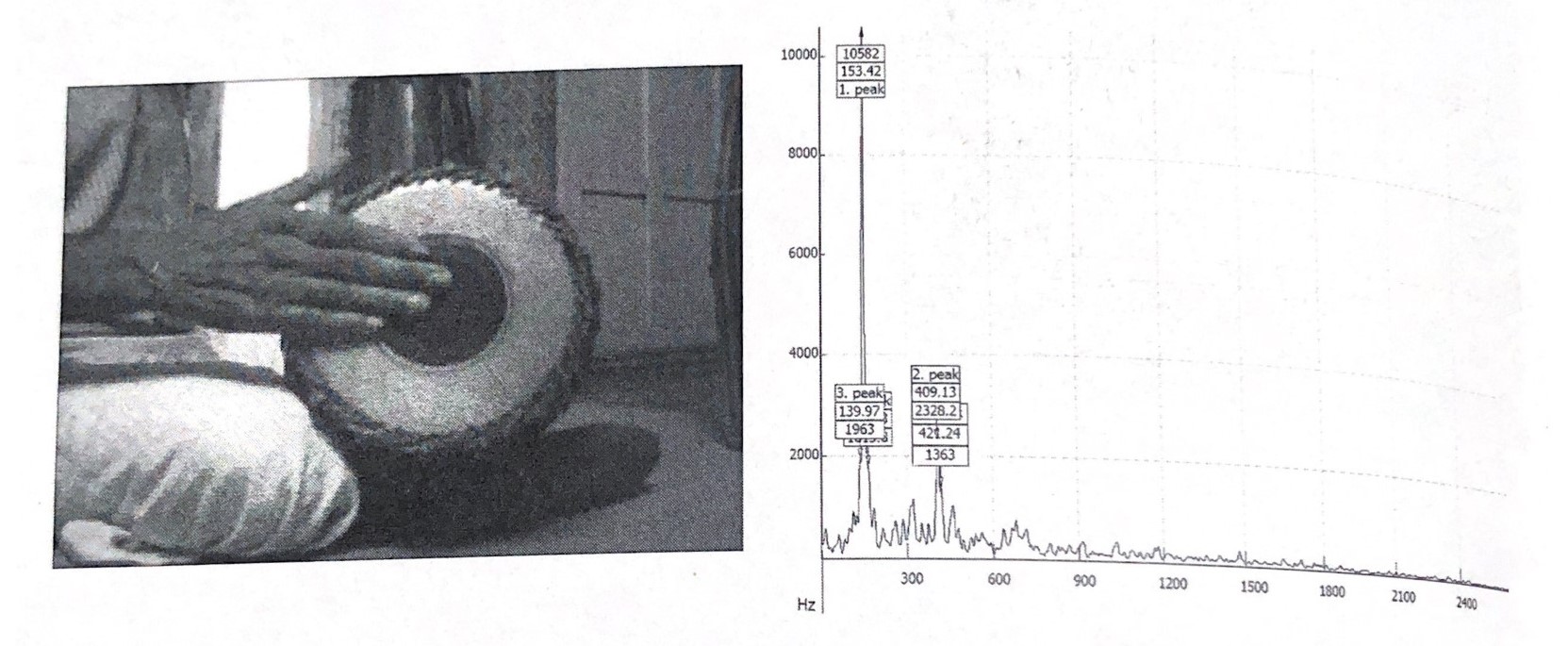}
    \caption{Dhi stroke with FFT spectrum}
    \label{fig:my_label}
\end{figure}

\textbf{Ta} \\
As seen in the measured FFT spectrum of Ta, it is a closed note that does not elicit harmonics, played on the center of the black patch. It has a sharp timbre due to its short sustenance time, which also allows it to be distinguished even when played multiple times a second. 
\begin{figure}[htb!]
    \centering
    \includegraphics[width=10cm]{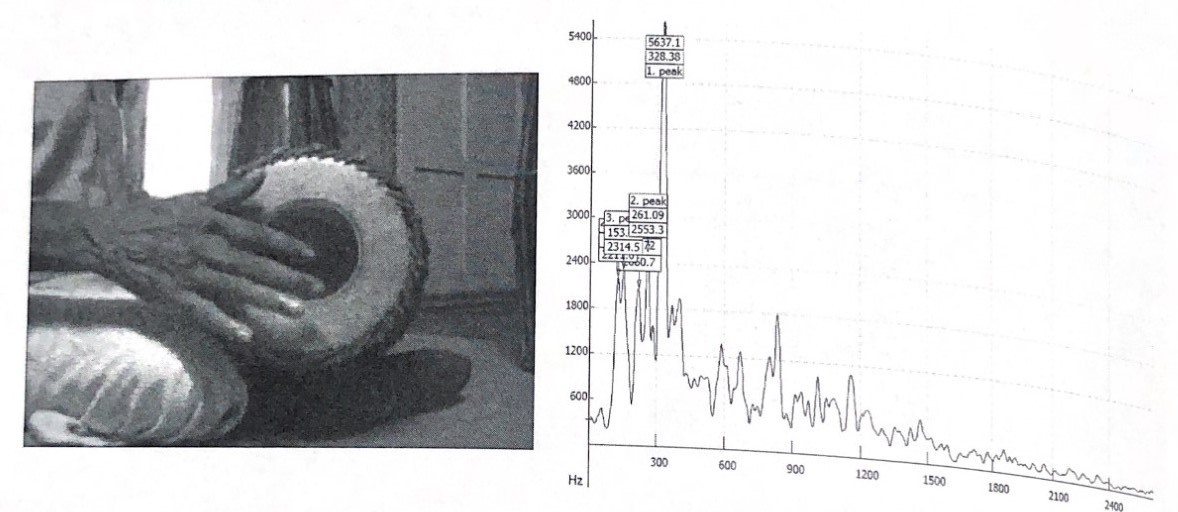}
    \caption{Ta stroke with FFT spectrum}
    \label{fig:my_label}
\end{figure}

\textbf{Thom} \\
Thom is an open base note played on the left side membrane. While popular opinion suggests that the left membrane does not produce harmonics, we see from the spectrum below that it is possible to extract harmonic peaks from thom as well. Since the left membrane is not centrally loaded and thus is anharmonic, one would not expect such harmonic peaks. However, these are able to be created because the artiste applies a paste of semolina on to the left membrane to mimic the role of the black patch of the right membrane. The application of this paste is adjusted so that the base tone of this membrane corresponds to the Pa note of the lower octave.
\begin{figure}[htb!]
    \centering
    \includegraphics[width=10cm]{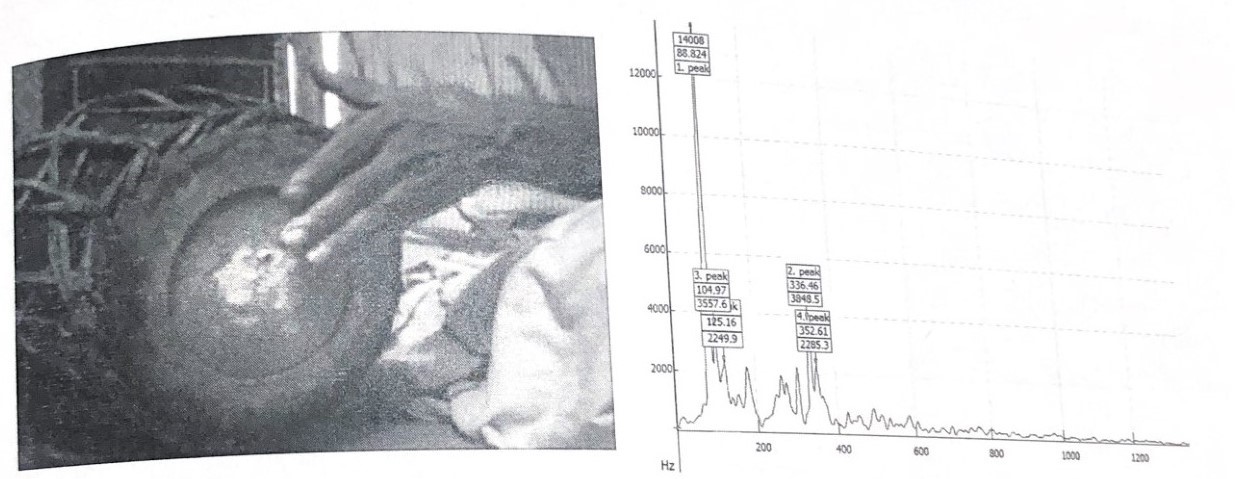}
    \caption{Thom stroke with FFT spectrum}
    \label{fig:my_label}
\end{figure}

\textbf{Gumkki} \\
Gumkki is another note that is harmonically unique to the mridangam (as seen in Table 6.3 of Musical Excellence and corresponding discussion). Dr. UKS has likened the naadham of gumkki to the 'cooing of the dove'. While it does display some harmonic characteristics as evidenced by the spectrum below, its primary use is to modulate the harmonicity of the thom strokes in-between which it is played.
\begin{figure}[htb!]
    \centering
    \includegraphics[width=10cm]{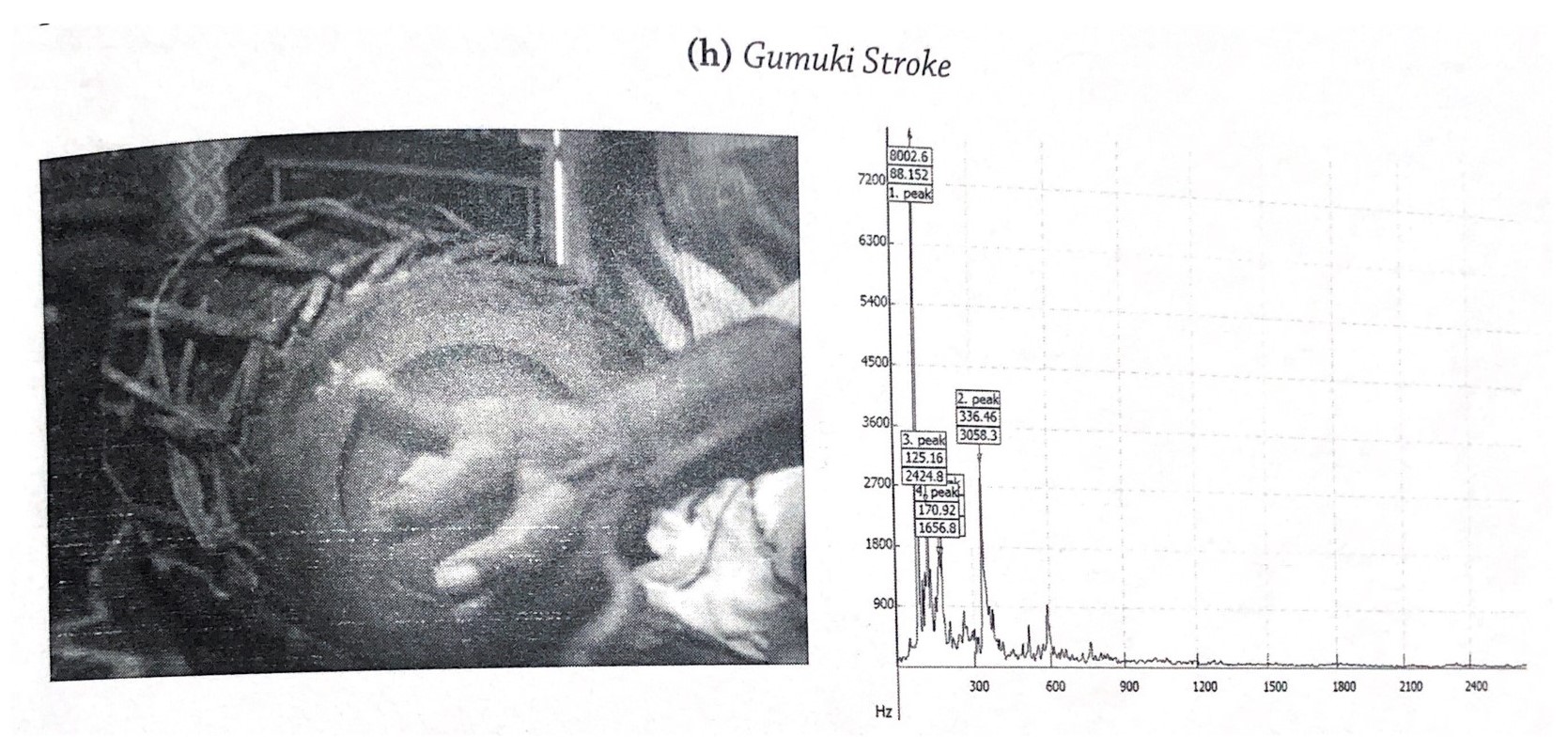}
    \caption{Gummki stroke with FFT spectrum}
    \label{fig:my_label}
\end{figure}

\subsection{Kucchi vs Thool Mridangam}
The issue of filling the annular space between membranes on the right side of the mridangam has been a source of intense discussion in the Carnatic music community for many decades. The reason behind this is likely due to the clearly different tonal qualities between 'kucchi' mridangams (annular space filled with sixteen equidistant strips of pine wood placed radially) and 'thool' mridangam (annular space filled uniformly with pieces of black patch material). In the end, it is more a matter of personal preference and artistic style as to which type is preferred, but it is important to understand the tonal differences between the two.

In Musical Excellence, the FFT spectrum of both 'thool' and 'kucchi' mridangams are presented and compared. It is found that both in terms of number of harmonic peaks observed and the intensity of the harmonic peaks, the dheem and nam strokes are more prominent in the thool mridangam, whereas the chappu and araichappu strokes are more prominent on the kucchi mridangam.

Additionally, Musical Excellence investigates the effect of inserting these annular materials, by looking at FFT spectra and time decay profiles before and after the insertion of 'kucchi'. It is seen that the insertion of 'kucchi' not only improves the sharpness of the harmonic peaks and hence the tonal quality, but also allows the note to be sustained for longer (seen through the time decay profile) when compared to membranes without annular filling. These annular materials can thus be used to adjust the sustenance of various notes on the mridangam. Musical Excellence further compares the role of these annular materials to that of the 'jeeva' in a tampura through similar FFT spectra. It is also interesting to see the sharpness of the harmonic peaks in a string instrument like the tampura, which is used for a purely melodic or tonal purpose, when compared to the mridangam.

While establishing the above facts about annular materials objectively is in itself a significant development, Musical Excellence goes further to propose a mechanism for the differences between thool and kucchi mridangams. To understand these interpretations, one first has to understand the shape of the modes excited by the different notes (fundamental/first harmonic for 'dheem, second for 'chappu' and third for 'nam'), seen in Figure 5.6 in Musical Excellence (also included here as Figure \ref{kucchithool}). Insertion of 'kucchi' prevents the whole membrane from moving as one, so modes that correspond to this kind of motion where either the membrane moves as a whole or the deformation is symmetric along the radius ('dheem' and 'nam') are suppressed. For these types of modes, we will find concentric circles on the membrane that are undergo no deformation - called nodal circles. On the other hand, insertion of 'kucchi' does not affect modes where there is no deformation of the membrane along the diameter (where the kucchis are placed) - also called a nodal diameter. Thus, the chappu and araichappu notes are not similarly suppressed in a kucchi mridangam, and are in fact enhanced. In the case of 'thool' mridangams, where particles are distributed throughout the annular space, the modes where the deformation is symmetric along the radius - which have nodal circles - ('dheem' and 'nam' ) are enhanced, whereas the modes with nodal diameters (chappu and araichappu) are suppressed. In Figure 6.5 (Chapter 6), Musical Excellence also shows beautiful images obtained using laser interferometry (an optical technique used to measure very small deformations), which show the shape of the different harmonic modes of the mridangam, which also help to illustrate this point. We also see, from the corresponding discussion, that a majority of these deformations occurs in the black patch region, highlighting its importance.
\begin{figure}[htb!]
    \centering
    \includegraphics[width=10cm]{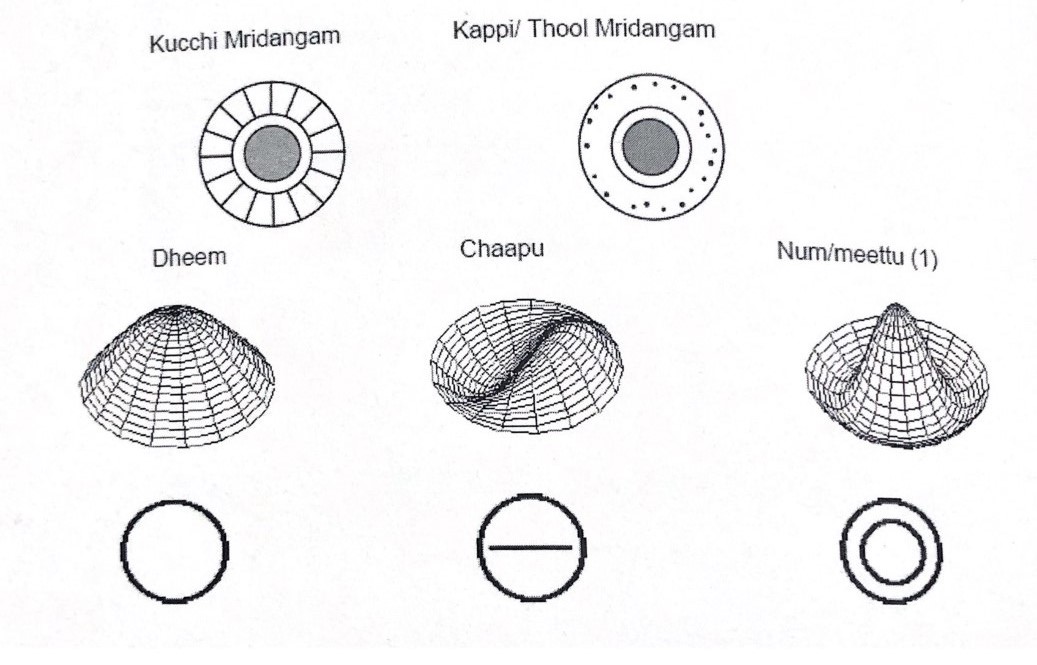}
    \caption{Shapes of first three modes of the mridangam, localized to the black patch}
    \label{kucchithool}
\end{figure}

\subsection{Cow-based membranes vs Goat-based membranes}
In addition to the annular filling material, the material used for the right membrane as the kottuthattu represents an important parameter for tonal quality. Traditionally, these membranes are made out of either goat or cow skin, as mentioned previously. Upon comparing FFT spectra, Musical Excellence finds that goat-based membranes seem to have sharper and higher intensity harmonic peaks, especially for the chappu stroke. Further study of these goat-based membranes using FFT spectra and time decay profiles reveals that thinner membranes seem to have better tonal quality and sustain of notes, with the chappu note sustaining without much loss of intensity for more than 1 second in thinned membranes. Thus, when comparing membranes across different materials or when proposing new materials to make these membranes, thickness of the membrane is an important parameter, and it could be possible that - up to a certain extent - thinner membranes allow for better tonal quality.
\subsection{Building on Raman's work}
As explained previously, Dr. CV Raman's work \cite{Raman1920,Raman1922,Raman1935} established the grounds for considering the mridangam as a harmonic instrument, unlike Western percussion instruments. Following his work and subsequent investigations, theoretical work by Malu and Siddharthan \cite{malu2000} has established the theoretical framework for this, and shown that the central loading in the form of the black patch is responsible for eliciting these harmonics, by considering a model with where the density (within the black patch region) varies as a function of the radial distance from the centre. With this model, they further predict that the frequency of the fundamental mode is shifted by 7 per cent when compared to the rest of the harmonics, which are in integer ratios as desired.

At the end of Chapter 5, Musical Excellence compares the results of their study using strokes played by Dr. UKS to this theoretical model, and find remarkable agreement in the frequency ratios of all the harmonics, including the seven percent shift in the fundamental. This further validates the continuous central loading model for the mridangam, and so, one may be able to use such a model to predict further musical outcomes (like the effect of density variations in the black patch, for example). Further, in the same section, Musical Excellence examines the time decay of various harmonics excited by the 'chappu' stroke, to investigate it's polytonic nature. It is indeed found that the second harmonic is the most prominent, and decays the slowest (longest sustaining). However, it is also found that the chappu stroke has a rich polytonic nature just after the instance of impact, with the third harmonic even exceeding the second harmonic in magnitude over a short time interval. The higher harmonics (including the third) decay faster than the second harmonic however, resulting in a transition to a fairly monotone sustained note after the first fraction of a second, which when tuned to the 'Sa' of the singer, produces a very pleasing effect.

Finally, in Figure 5.13, Musical Excellence illustrates the importance of the artiste's expertise in creating these harmonics and other tonal qualities, by comparing the FFT spectra of the chappu stroke played by two different artists. The first one shows clear and sharp peaks at the harmonic frequencies, which rise far above the background signals at other anharmonic frequencies, while the second shows much broader peaks which are not as clearly discernable since they are combined with background signals of comparable magnitude at the anharmonic frequencies. Hopefully, having read thus far, the reader can make a judgement on which is expected to elicit more 'naadham'!
\subsection{Chapter Summary}
This chapter of Musical Excellence thus uses the tools of Fourier Analysis and time decay profiles to uncover the unique tonal signatures of various strokes played on the mridangam and builds on the work of the previous chapter in analysing the role of various material choices like cow or goat based parchments, kucchi or thool etc, and other important factors like the skill of the artiste on the tonal signatures produced by various strokes like dheem and chappu. Further, it builds on the work of Dr. CV Raman by comparing various models of central loading through the black patch to experimental data obtained from advanced scientific instruments, thus giving further insight into the source of harmonics in this instrument. 
\newpage
\section{Towards Musical Excellence}
The main thrust of Musical Excellence is to pioneer the use of scientific tools in understanding the creation of naadham in the mridangam. To this end, the authors have looked at each of the various factors involved - including materials, construction, tonal properties and signatures and the skill of the artiste. There have been insights gained from each of these perspectives about the possible role of these different factors in creating the subjective experience of naadham. For example, we have seen how the addition of subsequent layers of black patch changes the frequencies of the strokes and the ratio between them, until it stabilizes to a fixed harmonic value. We have also seen how the insertion of kucchi creates sharper harmonic peaks for chappu and araichappu strokes, and discussed the possible mechanism behind this. The differences in tonal quality between thool and kucchi mridangams, and goat and cow based membranes have also been quantified, and the importance of the artiste's expertise in producing these tonal outcomes has also been qualitatively observed. The famed shifted fundamental has also been clearly demonstrated, as has the rest of the harmonic sequence - validating the model of Malu and Siddharthan \cite{malu2000}. These are a few of the research highlights of Musical Excellence, scattered throughout the pages, in between elucidating explanations and interesting conversations.

In Chapter 6, Musical Excellence synthesizes these findings and puts them together with a few more interesting ones - including a comparison of the harmonic nature of the mridangam and tabla - to present a full picture of the understanding that has been gained from this daunting undertaking. In addition to this, the authors use this understanding to propose solutions to the problems of standardizing the design of the mridangam, and finding alternative and sustainable materials for construction of the mridangam. Chapter 7 of Musical Excellence gives us a brief glimpse into this future, through a discussion of the SRI (Synthetic Rhythm Indian) Mridangam created by Dr. K. Varadarajan. 
\subsection{Characterizing Mridangams - Spectral ratios and time decay}
As mentioned earlier in this review, Musical Excellence establishes ratios between the frequencies of various strokes as characteristic of the mridangam, within error bounds: dheem to non-observable fundamental (1.07 $\mp$ 0.05), dheem to chappu (0.534 $\mp$ 0.005) and nam to chappu (1.5 $\mp$ 0.012 - (i.e) third to second harmonic). To find additional parameters of characterization, the authors further examine the time decay of the intensity of the dheem stroke in Figure 6.9 and the corresponding discussion and find a characteristic first-order decay rate constant of 2.02 s\textsuperscript{-1}, in addition to certain additional fine features like a beating pattern which could indicate that the dheem stroke profile is actually a combination of two closely spaced frequencies whose origin could well be related to the unique coupling between the two membranes of the mridangam which share a single resonating column. This bears further investigation, and is an interesting direction of future research.
\subsection{A comparison of Mridangam and Tabla - A case of resonating columns}
The tabla, in many ways, is the counterpart of the Mridangam in Hindustani music - due to its similar structure of membranes with a black patch, and similar role in concerts. The major point of difference between the two is that in the case of the table, the two membranes are contained in independent resonating columns which are only coupled through the air between and around them, whereas for the mridangam, both membranes share a common resonating column and are thus more strongly coupled. Comparing the tonal properties of these two instruments allows us to discern whether this coupling between membranes is important to producing harmonics, a point that was alluded to from the time of Dr. CV Raman. Upon obtaining FFT spectra for both Mridangam and Tabla tuned to various frequencies, the authors of Musical Excellence find that they are able to create a harmonic spectra in the tabla as well, indicating that the coupling between the membranes through the resonating column is not crucial to producing harmonics in these drum instruments, and further highlights the role of the black patch in producing said harmonics. Further, the authors are able to find notes played on the tabla that correlate to those on the mridangam in terms of their frequency ratios to the unobserved fundamental. 

However, there are a couple of interesting points of difference as well. There appears to be no equivalent note on the tabla for the Gumkki or Araichappu. Also, the ratio of the frequency of Dhi (the closest note to dheem in tabla) to fundamental is 1.12, which is different from the characteristic 1.07 for mridangam. This also means that the other characteristic ratio we discussed (dheem to chappu) would also be different for the table (In the case of the tabla, the 'Tin' note corresponds exactly to the second harmonic or chappu for the mridangam), although the 'Na' to 'Tin' ratio corresponds exactly to the Nam to Chappu ratio - reflected in the harmonic nature of the modes beyond the fundamental for both instruments. Therefore, from this method of characterization, Musical Excellence finds that the mridangam and tabla produce distinct tonal outcomes - even though they are similar in construction and production of harmonics. 

\subsection{Search for the fundamental mode - odd-even hypothesis}
It is indeed an amazing fact that the shifted fundamental seen when playing the dheem stroke exactly corresponds to the Rishabham or Ri note in Carnatic music. In Chapter 6, Musical Excellence explores a hypothesis where the Dheem note (which excites odd harmonics with a shifted fundamental) and the Chappu note (which excites even harmonic) correspond to two independent harmonic series - one corresponding to exciting axial modes and the other azimuthal modes. It is seen that the fundamental of the latter is suppressed with loading the black patch region, whereas the former remains prominent. Thus, it is conceivable that the makers of the mridangam cleverly juxtaposed these two independent sequences by accordingly adjusting the loading of the black patch such that the fundamental of the odd sequence corresponds to the Ri note if we fix the fundamental of the even note as Sa. This process - quantified by the oscillations in dheem to chappu frequency ratio - is beautifully seen in Figure 4.19 of Musical Excellence.

\subsection{Alternative materials for resonating column, membranes, black patches and insertion materials}
Using the criteria of SRC discussed earlier, FRP (Fibre Reinforced Plastic) and glass are presented as a possible alternative to wood for the resonating column of the mridangam. These materials would render the mridangam much lighter, and also allow for much smoother interiors of the resonating column - which is expected to influence tonal quality. 

As for the membranes, Musical Excellence discusses replacement of animal-based membranes by Mylar and Remo FybreSkin and motivates these suggestions by comparing the viscoelastic properties discussed earlier. It is found that these alternatives can reach comparable ratios of elastic and viscous deformations to the animal-based membranes, although the plastic component is slightly higher. Additionally, these alternative membranes do not have the same anisotropy in material properties along different directions as observed in the animal-based parchments, since they are not dependant on fibre arrangements and density. This also eliminates the non-uniformity in skin between different animals and parts of the same animal, found in animal-based parchments. 

\begin{figure}[htb!]
    \centering
    \includegraphics[width=10cm]{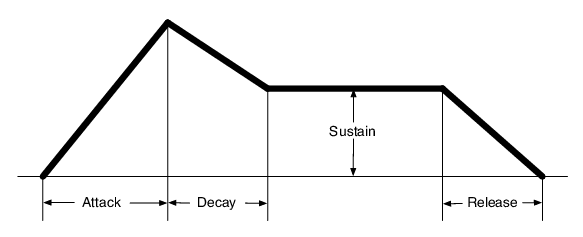}
    \caption{Schematic of an ASDR envelope. Image credits:\cite{ASDR} }
    \label{ASDR}
\end{figure}

For materials which fill the annular space, the proposed alternatives are divided into two main categories - materials that are distributed through out the annular space (thool-like) and materials that are inserted radially with a uniform separation (kucchi-like). Since the primary role of the annular material seems to be to arrest or amplify different vibrations and harmonics (As mentioned in Section 5.1.2 of this review, radial insertions enhance modes with nodal diameters (i.e) Chappu and Araichappu whereas distributed insertions enhance modes with nodal circles (i.e) Dheem and Nam ) , analysis is done using ASDR profiles - a common type of time decay profile (which we discussed in general before). ASDR stands for Attack, Decay, Sustain and Release of the musical signal - with Attack referring to the initial run-up from zero to highest intensity, Decay referring to the subsequent run-down to the level of intensity that is sustained, Sustain referring to the level of intensity maintained during the majority of the sound's duration and Release referring to the final run-down to zero intensity. A schematic of an ASDR profile is shown in Figure \ref{ASDR}.

For the thool replacements, steel balls of various uniform diameters are distributed throughout the annular space. For the kucchi replacement, Telfon strips of various uniform thicknesses are placed radially in the annular space, as metallic strips are found to add additional noise to the spectrum due to their own timbre. The authors are able to generate comparable ASDR profiles and FFT spectra for the various strokes using these alternatives, which also have the additional advantage of being of uniform thickness and material properties, which could potentially reduce tonal imbalances. In the case of radial insertions, it is found that thinner materials which have no metallic timbre of their own function well as dampeners.

The black patch is a crucial element of the unique tone of the mridangam, and thus finding potential alternatives for the Iron oxide based black patch material and the cooked rice adhesives is an important problem. Musical Excellence examines various adhesive materials through their compressive strength and found that polymers like Carboxyl Methyl Cellulose (CMC) and Poly Vinyl Pyrollidone (PVP) performed better than cooked rice, and thus could be potential replacements which prolong the life of the black patch. As for the black patch material itself, replacement of the iron oxide with zirconium salts to produce a white patch was experimented with. Upon constructing an instrument with such a white patch, it was found that significant deviations of the dheem stroke from the expected shifted fundamental was observed, even though the harmonic nature of the higher harmonics was retained. This shows that the material properties of the iron oxide black patch are important to the tonal properties of the mridangam, and this is a promising direction of further investigations.

Additionally, Musical Excellence also considers the effect of treating the parchments with different tanning agents through comparing their porosity and densities, and their viscoelastic properties. Even though it was found that treating with tanning agents did not significantly affect tensile and other physical properties, the tonal properties were found to be inferior to the untanned parchments.
\subsection{Alternate Mridangam Designs}
To round off Chapter 6, Musical Excellence presents several interesting designs for transport-friendly mridangams and evaluates the tonal properties of these designs through constructed prototypes. We saw in earlier sections that the nature of the harmonics produced by the mridangam are crucially dependent on the black patch, but not significantly affected by the resonating column. Musical Excellence uses this to its advantage by developing alternative designs that modify the resonating column to make it more transport friendly, thus hoping to retain the tonal qualities. These are then characterized using FFT profiles and time decay of strokes played by Dr. UKS, similar to analysis done in previous chapters. Design I considers a mridangam that is split into four parts, which can then be reassembled easily. The buffalo hide straps were limited to the parts on the extremities, while the central wooden resonating column is bare. The tonal properties were reproduced sufficiently well, with surprisingly sharp harmonic peaks. However, the challenges in this design are the difficulty of fabrication and the still insufficient tonality from the artiste's point of view. Design II considers a foam based resonator which can be attached on to and removed from the mridangam body as required. However, the tonal quality of this design was inferior as evidenced by broader harmonic peaks across most strokes played. ASDR profiles also reveal that the sustenance is insufficient. Design III is similar to Design I, but made so that outer parts go inside the inner resonating column parts, making it further suitable for transport. The tonal qualities and harmonic peaks were similar to Design I with similar drawbacks. Design IV considers a telescopic design, and Design V considers a 3-in-1 mridangam with multiple heads with varying tonal properties being placed inside the same body of the instrument. This design too produced sufficiently sharp harmonic peaks, but the tonal quality was not sufficient from the artiste's point of view.

Thus, while the fundamental features like the shifted fundamental and the dheem to chappu frequency ratios are demonstrated in these prototypes, their tonal quality remains to be rigorously evaluated by artistes and scientists alike. However, Musical Excellence does give us a lot of the tools necessary for this task. 

These alternative materials, designs and the analysis techniques used to study them enable Musical Excellence to take a big step towards accomplishing its stated goals of standardizing materials and fabrication of the mridangam, develop sustainable alternatives and improve tonal outcomes. Along the way, Musical Excellence has contributed greatly to understanding the tonal qualities of this unique instrument.
\begin{figure}[htb!]
    \centering
    \includegraphics[width=6cm]{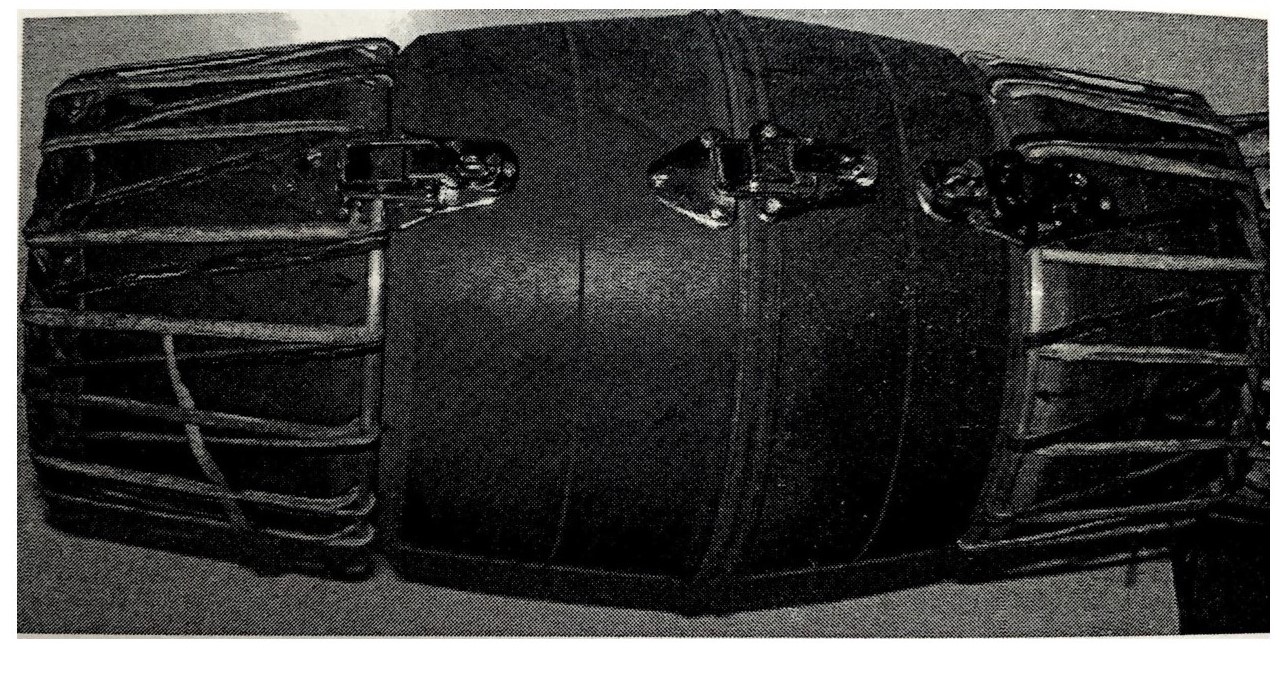}
    \includegraphics[width=6cm]{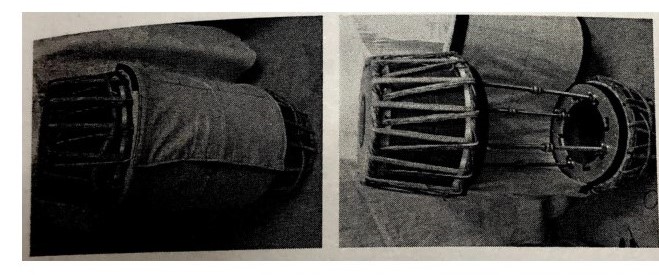}
    \includegraphics[width=6cm]{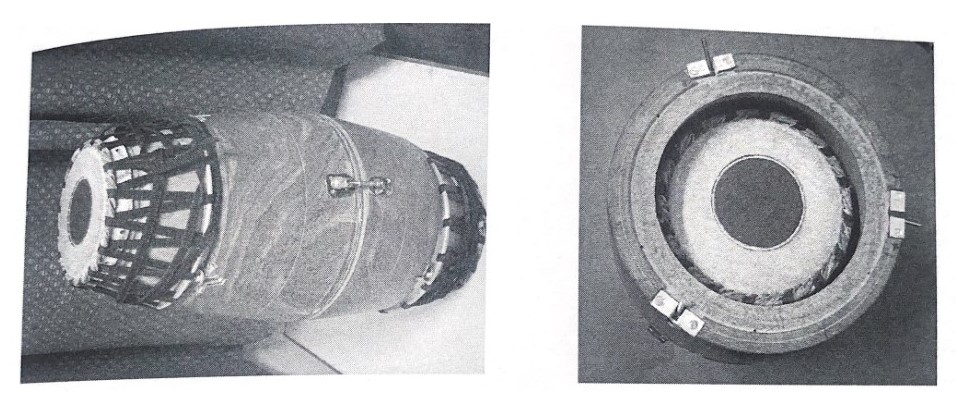}
    \includegraphics[width=6cm]{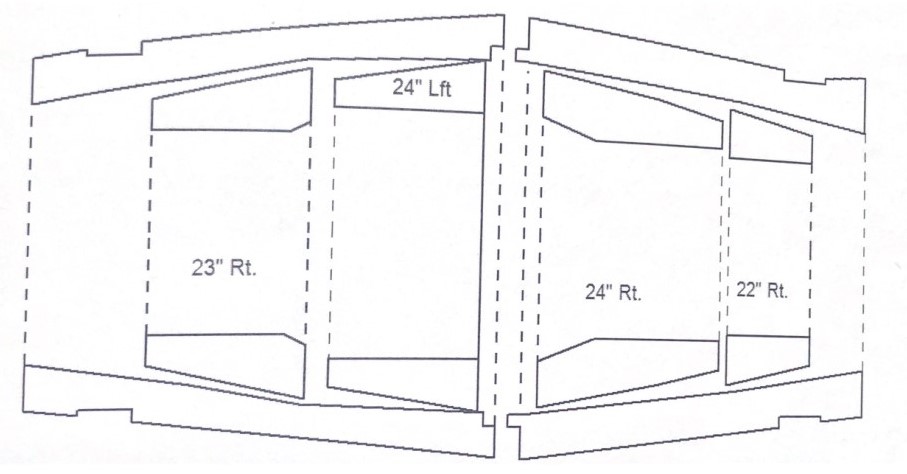}
    \caption{Prototypes of various alternate mridangam designs discussed in Musical Excellence}
    \label{kucchithool}
\end{figure}

\subsection{SRI Mridangam - a vision to the future}
The final chapter of Musical Excellence concerns the SRI Mridangam, a design established by Dr. Varadarangan, an excellent first step towards the standardization of synthetic mridangams. Taking inspiration from Western drums, the drum head is made out of polyester film and the black patch out of a special type of rubber that is chemically bonded to the film. The wooden shell is replaced with fibre glass, reducing the weight of the instrument significantly. The drum head was attached to this shell with metal clamps tensioned with nuts and bolts. 

The overall final design was developed iteratively, and the whole process is documented in an interesting way in the final chapter of Musical Excellence, written by Dr. Varadarangan. The mridangam went through extensive testing to ensure that it withstands wear and tear of daily use without damage, and the final design is optimized to ensure such robustness. Further testing was done to compare the tonal quality of the SRI mridangam to the traditional mridangam across the various strokes played on the right drum head, and it was found to have comparable tonal properties for each of the strokes. 

The main advantages of the SRI mridangam are that it is environmentally and animal friendly, light weight and easily transportable, has a consistent tonal quality that is robust against external conditions, a wide tuning range, and easier to maintain. While it is a great feat that a comparable tonal quality to the traditional mridangam has been achieved in terms of FFT spectra and time-decay profiles, it will be a test of both the instrument design of the SRI mridangam and these characterization methods which we have extensively discussed here, to see whether these results translate into a similar subjective listening experience as the traditional mridangam.


\subsection{Chapter Summary}
The final chapters of Musical Excellence thus looks towards the future in improving our understanding of this wondrous instrument by presenting hypotheses and interpretations of the source of its unique tonal properties, especially the shifted fundamental. Further insight is also gained by comparing the tonal signatures to that of the tabla, which also allows one to understand the role of the common resonating column in the case of the mridangam versus the independent resonating columns in the case of the tabla. The second major objective accomplished by these chapters is synthesizing the knowledge gained from the comprehensive analysis done to propose new designs for the mridangam that satisfy modern requirements of sustainability, animal-friendliness and transportability, while retaining the essential tonal qualities. Of course these designs and the materials involved will have to be improved upon and optimized, but Musical Excellence offers both a great starting point for new designs and solid framework for improving these designs using objective scientific analysis that is augmented by artistic insight.
\newpage
\section{Epilogue}
Scientific analysis of musical instruments is a field which has yet to reach maturity both in terms of clearly establishing relevant scientific questions, and the techniques used for such an analysis. While important scientific breakthroughs regarding the structure and the production of musical notes have been made especially in the case of the violin, a unifying framework of questioning and analysis is still lacking. The fact that music is a subjective human experience means that it is exceedingly difficult to establish what questions can be asked that can be evaluated scientifically. These factors are further magnified in the case of Indian instruments since the materials used and processes of fabrication, which are millennia old, are typically only known by generations of skilled artisans who learn these through an experiential and oral tradition. Thus, the task set forth for Musical Excellence is indeed a daunting challenge. 

The main factor which has allowed Musical Excellence to surpass this challenge is the close collaboration between a legendary artist in Dr. UKS, eminent scientists in Dr. Ramasami and Dr. Naresh, and a skilled artisan in Sri Johnson. Through this collaboration, the authors of Musical Excellence first focus on establishing the current level of understanding about the construction and music produced by the mridangam, and laying down the important research questions that can potentially be addressed through scientific analysis. The authors then go on to establish the current level of scientific understanding and the scientific tools that could be used to address the above questions. Then, following rigorous scientific investigation, Musical Excellence is able to establish important results that improve our understanding of both the construction and materials of the mridangam and the unique nature of musical notes produced - like the observation of the shifted fundamental excited by the dheem stroke, the role of various annular materials like kucchi and thool and the importance of layer-by-layer application of the black patch - thus validating and moving beyond the work of Dr. CV Raman. Finally, the authors also use the improved understanding that has been gained to propose new designs and standardization techniques for the mridangam, laying the foundations for a future in which the mridangam is both travel-friendly and made from sustainable materials in a standardized fashion, yet preserving the signature naadham of the traditional mridangam. 

Musical Excellence of Mridangam by Dr. Umayalpuram K Sivaraman, Dr. T Ramasami and Dr. M D Naresh is thus an important body of research in the field of Carnatic music and even music in general, and is a shining example of how scientific research can be done in a field that is built on human subjectivity and whose subject is shrouded in ancient myths and mysteries.

\end{document}